\pdfoutput=1 
\documentclass[preprint,authoryear,12pt]{elsarticle}

\usepackage{graphicx}

\usepackage{amssymb}

\usepackage{hyperref}

\journal{Advances in Space Research}

\def\cp{C$^+$}
\def\hp{H$^+$}
\def\op{O$^+$}
\def\hh{H$_2$}
\def\hho{H$_2$O}
\def\hhoe{H$_2^{18}$O}
\def\hhos{H$_2^{17}$O}
\def\oo{O$_2$}
\def\ohp{OH$^+$}
\def\hhop{H$_2$O$^+$}
\def\hhhop{H$_3$O$^+$}
\def\meth{CH$_3$OH}
\def\chp{CH$^+$}
\def\ammo{NH$_3$}

\def\mic{$\mu$m}
\def\gtsim{{_>\atop{^\sim}}}
\def\ltsim{{_<\atop{^\sim}}}
\def\scm{cm$^{-2}$}
\def\ccm{cm$^{-3}$}
\def\pow#1#2{#1$\times$10$^{#2}$}
\def\dv{$\Delta${\it V}}
\def\kms{km\,s$^{-1}$}
\def\vlsr{$V_{\rm lsr}$}

\def\new#1{{#1}}

\begin{document}

\begin{frontmatter}



\title{The first results from the Herschel-HIFI mission\tnoteref{footnote1}}
\tnotetext[footnote1]{\textit{Herschel} is an ESA space observatory with science instruments provided
by European-led Principal Investigator consortia and with important participation from NASA}


\author{Floris van der Tak\corref{cor}\fnref{footnote2}}
\address{SRON Netherlands Institute for Space Research, Landleven 12, 9747 AD Groningen, The Netherlands}
\cortext[cor]{Corresponding author}
\fntext[footnote2]{Also affiliated with Kapteyn Astronomical Institute, University of Groningen, The Netherlands}
\ead{vdtak@sron.nl}




\begin{abstract}
This paper contains a summary of the results from the first years of observations with the HIFI instrument onboard ESA's Herschel space observatory. 
The paper starts with a summary of the goals and possibilities of \new{far-infrared and} submillimeter astronomy, the limitations of the Earth's atmosphere, and the scientific scope of the Herschel-HIFI mission. 
\new{The presentation of science results from the mission follows the life cycle of gas in galaxies as grouped into five themes: Structure of the interstellar medium, First steps in interstellar chemistry, Formation of stars and planets, Solar system results and Evolved stellar envelopes.}
The HIFI observations paint a picture where the interstellar medium in galaxies has a mixed, rather than a layered structure; the same conclusion may hold for protoplanetary disks. 
In addition, the HIFI data show that exchange of matter between \new{comets} and asteroids with planets and moons plays a large role.
The paper concludes with an outlook to future instrumentation in the far-infrared and submillimeter wavelength ranges.
\end{abstract}


\end{frontmatter}

\parindent=0.5 cm

\newpage

\begin{figure}[tb]
\centering
\includegraphics[width=0.7\textwidth,angle=0]{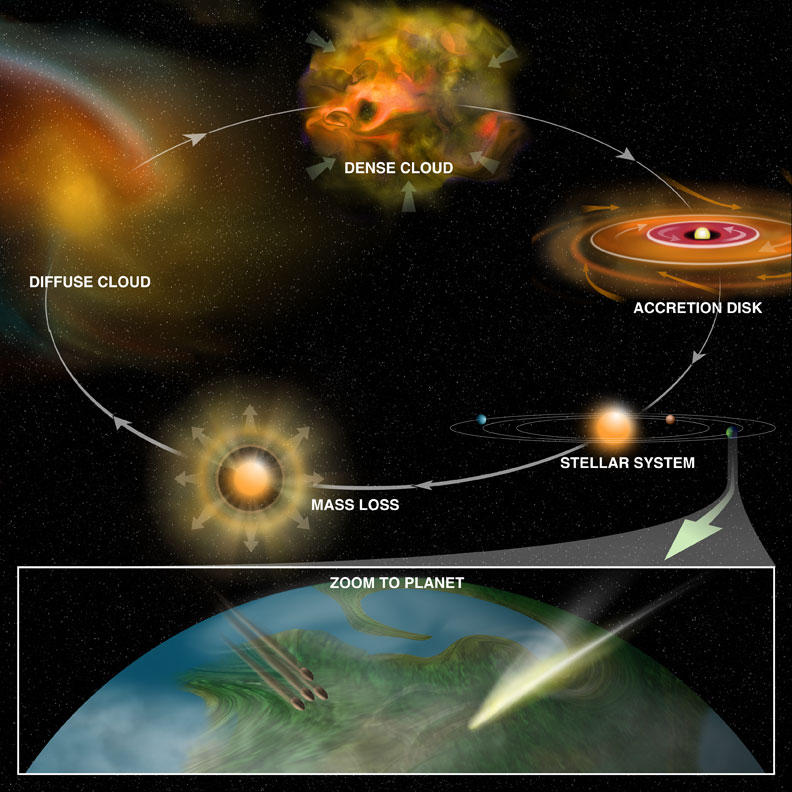}
\caption{The cycle of gas and dust in galaxies: from interstellar clouds to stars and planets. Credit: Bill Saxton (NRAO/AUI/NSF).}
\label{f:cycle}
\end{figure}

\section{Introduction}
\label{s:intro}

The Herschel space observatory is a facility to study \new{celestial objects in} far-infrared light, which is the third most common kind of radiation in the Universe\footnote{See {\tt http://astro.ucla.edu/$\sim$wright/CIBR/}}. As measured by energy density, the most common type of radiation are photons from the Cosmic Microwave Background, which were emitted in the early Universe when protons and electrons first (re)combined to form hydrogen atoms; studying this type of light is studying our past. The second most common type of radiation ($\approx$6\% of the CMB intensity) is optical and near-infrared light, which is emitted by the current generation of stars; studying this light is studying our present. At $\approx$3\% of the CMB intensity, far-infrared light is about half as common \new{as} optical light, and at least as important: this light is emitted by cold clouds of gas and dust where new stars and planets are forming (Figure~\ref{f:cycle}). Studying far-infrared light from the local Universe is therefore studying our future.

\begin{figure}[tb]
\centering
\includegraphics[width=0.9\textwidth,angle=0]{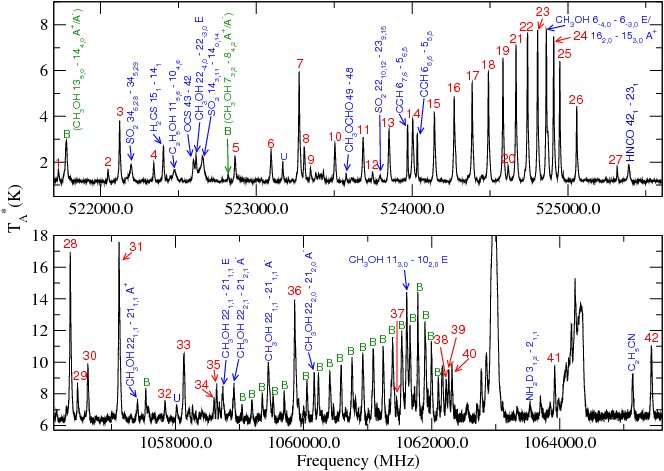}
\caption{The HIFI spectrum of the Orion-KL region shows a multitude of lines from which many physical and chemical properties of the gas can be inferred. \new{Red numbers denote isolated methanol transitions, blue text denotes transitions from other molecules and methanol transitions which are blended, green text and ÒBÓ denotes methanol lines which are blended with different parity states, and ÒUÓ denotes unidentified lines. The chemical compounds seen in this spectrum were known from ground-based observations, but these lines are seen with HIFI for the first time.} From Wang et al (2011).}
\label{f:orion}
\end{figure}

The far-infrared and submillimeter part of the spectrum is well suited to study the origin of galaxies, stars and planets, because gas and dust clouds with temperatures of 30--100\,K emit the bulk of their radiation in this range. In particular, continuum observations readily probe the mass and the temperature of the clouds. Here, the advantage over mid-infrared or shorter-wavelength observations is that the radiation is mostly optically thin, so that it traces the entire volume of the clouds rather than just their surfaces. Furthermore, the large number of atomic fine structure and molecular rotational transitions spanning a wide range of excitation energies (from $\sim$1 to $\sim$1000\,K) and radiative decay rates \new{provides} a powerful tool to measure the densities, temperatures and masses of interstellar clouds (Figure~\ref{f:orion}). In addition, observations at high spectral resolution enable to study the kinematics of clouds in detail, polarization measurements give information about magnetic fields, and the chemical composition of clouds contains signatures of otherwise hidden (energetic) radiation fields. The goal of far-infrared and submillimeter astronomy is therefore a basic understanding of the physics and chemistry of interstellar clouds, star-forming regions, protoplanetary disks, the envelopes of evolved stars, planetary atmospheres, active galactic nuclei, and starburst galaxies.

Ground-based telescopes can only partly reach the above goals, because the Earth's atmosphere is opaque for a large fraction of the submillimeter range and all of the far-infrared. Due to pressure broadening, absorption by atmospheric gases not only blocks our view of abundant atmospheric constituents such as \hho\ and \oo, but also of many other species and of entire frequency ranges, including the entire far-infrared at frequencies above $\sim$1000\,GHz, including the fine structure lines of \cp\ and O at 1901 and 4745\,GHz which require airborne observatories such as SOFIA. Below $\sim$1000\,GHz, the transmission at some frequencies is so low that ground-based observations are limited to bright sources and small regions on the sky; an important example are the fine structure lines of C at 492 and 809\,GHz, which can only  under favourable conditions be observed from high-altitude observatories such as Mauna Kea (JCMT, CSO, SMA) and Chajnantor (APEX, ALMA).

\new{Following successful missions such as ISO and KAO,} the Heterodyne Instrument for the Far Infrared (HIFI) onboard ESA's Herschel space observatory has been 
designed to give astronomers near-complete coverage of the submillimeter waveband. 
The orbit of Herschel around the L2 point and the passively cooled mirror provide superior sensitivity and stability compared with ground-based observatories.
The spectral coverage of HIFI ranges from 480 to 1250 GHz in five bands and from 1410 to 1910 GHz in two additional bands, or in other words from the fine structure lines of C to that of \cp\ with a gap around the p-H$_2$D$^+$ ground state line. By mixing the sky signal with a locally generated frequency standard, HIFI achieves a spectral resolution of $\approx$0.1\,MHz or a resolving power of $\sim$10$^7$. The angular resolution is set by the diffraction limit of Herschel's 3.5-m mirror and ranges from 39$''$ at the lowest to 13$''$ at the highest frequencies. The instrument measures a single beam on the sky, but maps can be made by scanning the telescope and combining the data afterward. See \citet{pilbratt2010} for a description of the Herschel telescope, \citet{degraauw2010} for a description of the HIFI instrument, and \citet{roelfsema2012} for its in-orbit calibration and performance.


This paper reviews the results from the first half of the Herschel-HIFI mission, grouped into five themes: \new{Structure of the interstellar medium (\S~\ref{s:ism}), First steps in interstellar chemistry (\S~\ref{s:chem}), Formation of stars and planets (\S~\ref{s:psf}), Solar system results (\S~\ref{s:solsys}), and Evolved stellar envelopes (\S~\ref{s:agb})}. Results from Herschel's other two instruments, PACS and SPIRE, will only be mentioned to provide context for HIFI results; space limitations prevent a more complete coverage of PACS and SPIRE results. Also, the science cases for each of the selected HIFI results will necessarily be brief. For a general introduction in the field of interstellar medium physics, see the book by \citet{draine2011}. The formation of stars is discussed extensively in the book by \citet{palla-stahler}. An up-to-date comprehensive review of interstellar chemistry does not exist currently, but the proceedings of IAU Symposium 280 provide a good starting point. The envelopes of evolved stars are reviewed by \citet{habing1996}, while the book by \citet{depater-lissauer} covers Solar system science.


\section{Structure of the interstellar medium}
\label{s:ism}

The space between the stars is far from empty: about 10\% of the baryonic mass in galaxies is contained in the interstellar medium (ISM). This medium consists of several phases: hot and warm ionized gas, warm and cold neutral gas, and molecular clouds, which are in approximate pressure equilibrium. While the warm diffuse ionized and atomic phases occupy most of the ISM volume, the cold dense molecular phase contain most of the mass. In this latter phase, pressure equilibrium breaks down and self-gravity becomes important, which leads to gravitational collapse and the formation of new generations of stars and planets. The cold dense ISM of galaxies is a key science area for Herschel, because many questions about its nature can only be addressed with far-infrared observations which are impossible from the ground. In particular, HIFI observations are instrumental to study the physical conditions, the geometrical structure, and the chemical composition of the ISM, as detailed in the following subsections.

\subsection{A new population of interstellar clouds}
\label{ss:wdg}

\begin{figure}[tb]
\centering
\includegraphics[width=\textwidth,angle=0]{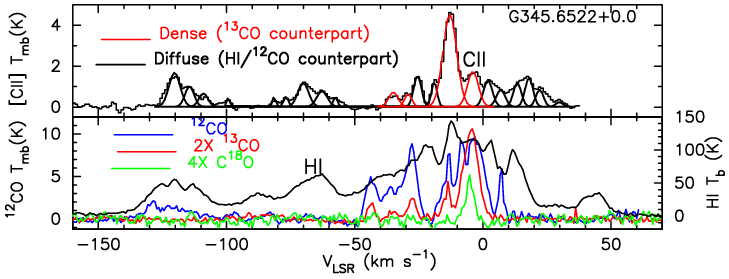}
\caption{Top panel: Spectrum of the \cp\ fine structure line \new{observed with HIFI} toward a region near the Galactic Center, with a decomposition into Gaussian components. Bottom panel: Spectra of the CO $J$=1--0 and \new{HI 21cm} lines toward the same position, showing that some clouds are mainly atomic, others mainly molecular, and others of mixed nature. From Langer et al (2010).}
\label{f:gotc+}
\end{figure}

A key objective of the HIFI mission are observations of the fine-structure line of \cp\ at 158\,\mic, which is the primary coolant of diffuse interstellar gas clouds. Measurements with the COBE, BICE and KAO observatories have revealed the large-scale distribution of \cp\ emission, but HIFI offers superior angular and spectral resolution which is necessary to disentangle close neighbours on the sky and along the line of sight. 

\new{Figure~\ref{f:gotc+} shows} the most surprising result of the \cp\ observations with HIFI: that significant \cp\ emission is observed from interstellar clouds where most hydrogen is in the form of \hh\ but carbon is not locked up in CO. This regime of physical conditions \new{($T \approx 100$\,K, $n_H \approx 300$\,\ccm, $A_V = 0.1 - 1.3$\,mag)} is referred to as "warm dark gas" \citep{langer2010}. 
The amount of gas under these conditions is much larger than previously assumed \new{from $\gamma$-ray data (e.g., \citealt{grenier2005})}: as much as $\approx$25\% of the total \hh\ in these \new{`warm dark'} clouds may be in \hh\ layers which are not traced by CO \citep{velusamy2010}.
\new{This fraction may be much higher in dwarf galaxies, where $L$[CII]/$L$(CO) ($\gtsim$10$^4$) is at least an order of magnitude greater than in the most metal-rich starburst galaxies \citep{madden2011}.}

The power of HIFI for understanding extragalactic \cp\ emission is nicely illustrated by \citet{mookerjea2011}, who present a study of the gas around the \hp\ region BCLMP 302 in the nearby spiral M33. The widths of the \new{[CII]} line profiles are found to be intermediate between those of the CO and \new{HI 21cm} lines, and the spatial correlation of \cp\ with both CO and H is found to be rather poor. The authors estimate that about $\approx$25\% of the \cp\ emission has an origin in ionized gas. 


\subsection{Thermal balance of dense interstellar clouds}
\label{ss:thermal}

The cooling of dense molecular clouds is a long-standing question of astrophysics. Models of pure gas-phase chemistry predict large abundances of \oo\ and \hho, which would make these molecules the major carriers of oxygen and major coolants of gas at densities $\gtsim$10$^4$\,\ccm\ \citep{goldsmith1978,bergin1995}. Thermal emission from these species can only be observed from space, and the first tests of those predictions were made by the SWAS and Odin missions around the year 2000. On the pc-sized scales of interstellar clouds, the measured \hho\ abundance was found to be orders of magnitude below the gas-phase predictions, probably due to depletion of oxygen on grain surfaces \citep{melnick2005}. The \oo\ molecule was only detected toward one position \citep{larsson2007}; this result and upper limits toward many other sources also indicate a low abundance.

The HIFI instrument is much better suited to search for interstellar \oo\ than SWAS and Odin for three reasons. Its lower system temperature implies a much better sensitivity, its larger frequency coverage enables observation of three transitions instead of just one, and its 3--10 \new{times} smaller beam size is better coupled to the expected sizes of the emitting regions. The result is the first clear detection of interstellar \oo\ toward Orion \citep{goldsmith2011} and the confirmation  of the Odin detection toward $\rho$~Oph \citep{liseau2011}. Upper limits have been obtained toward many other sources, which indicates that the formation of \oo\ is very sensitive to local conditions. Clearly, the role of \oo\ as oxygen carrier and gas coolant is very limited.

Like for \oo, HIFI is well suited to characterize the distribution and excitation of interstellar \hho, since many transitions (including the isotopologs \new{\hhoe,\hhos, and HDO}) lie within its frequency range. Many Herschel programs include observations of \hho, as a probe of physical conditions, kinematics, or chemistry. The basic result of these studies is that while \hho\ may locally reach high abundances, particularly in shocks (\S~\ref{ss:shc}), the bulk of interstellar gas contains very little \hho. The typical abundance of \hho\ in non-shocked gas is $\sim$10$^{-9}$, which is limited by freeze-out on grain surfaces in dense clouds, and by photodissociation at low densities (e.g., \citealt{vandertak2010}).  This result implies that in the general interstellar medium, \hho\ is not a major oxygen carrier, nor a major coolant, confirming earlier \new{results} from SWAS \citep{bergin2000}. 

\subsection{The transition between diffuse and dense interstellar clouds}
\label{ss:low-n}

In cold dense interstellar gas clouds, the bulk of the hydrogen is in the form of \hh, which is basically unobservable at low temperatures. The common proxy for \hh\ is the CO molecule, which is abundant and chemically stable, but which is prone to photodissociation at \new{$N$(\hh)$\ltsim$\pow{1.5}{21}\,\scm\ \citep{visser2009}}. A good alternative for this regime is HF, which is also chemically very stable. The reaction of F with \hh\ leading to HF is exothermic, so that HF is the main carrier of gas-phase fluorine, especially at low (column) densities. 

Observations with ISO have confirmed these predictions, although they only probe excited states of HF \citep{neufeld1997}. The HIFI instrument gives our first access to the rotational ground state of HF, and observations of widespread absorption in the $J$=1--0 line indicate an HF abundance of \pow{(1--2)}{-8} in diffuse clouds, close to the interstellar fluorine abundance \citep{sonnentrucker2010,monje2011}. Towards dense clouds, the inferred abundance is $\sim$100 \new{times} lower \citep{phillips2010}, suggesting that depletion on grain surfaces or excitation effects play a role. 

The large dipole moment of HF and the high frequency of its 1--0 line imply a rapid radiative decay rate, which explains why the line usually appears in absorption. Only in very dense environments such as the inner envelopes of late-type stars, the line appears in emisison \citep{agundez2011}. 
The only detection of HF emission from the Galactic interstellar medium so far is toward the Orion Bar \citep{vandertak2012:hf}, which is a surprise because the \hh\ density in this region is not high enough to excite the line. Instead, collisional excitation by electrons  appears to dominate, whereas non-thermal excitation mechanisms such as infrared pumping or chemical pumping appear unlikely \citep{vandertak2012:h3+}. Emission lines of HF thus appear to trace regions with a high electron density, caused by strong ultraviolet irradiation of dense molecular gas.

The high electron density in the Orion Bar may also apply to the active nucleus Mrk 231 where the SPIRE spectrum shows HF emission \citep{vanderwerf2010}. Other active galactic nuclei such as the Cloverleaf quasar\citep{monje2011clover} show HF in absorption, suggesting that these nuclei have low electron densities, \new{while yet others such as Arp 220 \citep{rangwala2011} show P~Cygni profiles, suggesting the presence of HF in a wind with a high electron density.}


\subsection{Mixing of atomic and molecular phases}
Models of the structure of interstellar gas clouds predict that at low densities, hydrogen is primarily in atomic form, while at higher densities, the molecular form predominates. It is therefore common practice to divide such clouds in two classes: atomic clouds and molecular clouds (e.g., \citealt{snow2006}). Observations of ionized water species with HIFI have however changed this picture.

The formation of \hho\ in cold gas starts with \op, reacting with \hh\ to first form \ohp, then \hhop, and then \hhhop, which recombines with an electron to form \hho. These reactions have no barriers and proceed fast, so that no significant amounts of the intermediate products \ohp\ and \hhop\ are expected if all hydrogen is in the form of \hh. 

Observations with HIFI have revealed large column densities of both \ohp\ and \hhop, which is inconsistent with the above predictions \citep{ossenkopf2010,gerin2010}. Before the Herschel mission, \hhop\ was only known in comets \citep{herzberg1974} and only a tentative detection of \ohp\ with ISO was reported toward by \citet{gonzalez2008}.  The implication of widespread interstellar \ohp\ and \hhop\ is that `mixed' clouds appear to be common, where hydrogen is partly H and partly \hh\ \citep{neufeld2010:hno+}. The ratio of \hhop\ to \hho\ is observed to vary, probably due to variations in the density of the clouds \citep{wyrowski2010}.

\begin{figure}[tb]
\centering
\includegraphics[width=0.5\textwidth,angle=0]{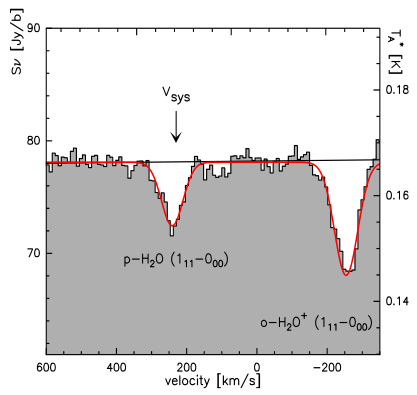}
\caption{Spectrum of \hho\ and \hhop\ lines toward the active nucleus of M82. The high \hhop/\hho\ ratio indicates fast photodissociation of \hho\ by the strong UV radiation field from the starburst. From Wei{\ss} et al (2010).}
\label{f:weiss}
\end{figure}

HIFI has also observed ionized water in nearby starburst galaxies such as M82, NGC 253, and NGC 4945 (Figure~\ref{f:weiss}). The observed column density of \hhop\ in M82 is almost as high as that of \hho\ \citep{weiss2010}, probably because \hho\ suffers from rapid photodissociation by ultraviolet radiation from the young stellar population, while \hhop\ is much more robust against ultraviolet light. Even more extreme conditions occur in the active galactic nucleus of Mrk 231, where lines of \ohp\ and \hhop\ appear in emission \citep{vanderwerf2010}. This unusual phenomenon may be the result of rapid formation via X-rays or cosmic rays, along with strong excitation by electron collisions.

\section{First steps in interstellar chemistry}
\label{s:chem}

The chemical composition of interstellar clouds and star-forming regions gives valuable clues to physical processes which are not directly visible. Spectral surveys at (sub)millimeter wavelengths from the ground have shown some of this potential but are limited to atmospheric windows. The HIFI instrument gives us the first full view of the molecular line spectra in the 500--2000\,GHz range. Besides the major oxygen carriers \hho\ and \oo\ discussed above, the HIFI range includes several key species to improve our understanding of the physics of interstellar clouds. This section reviews the areas where HIFI has already contributed to this understanding.

\subsection{Turbulent chemistry}
\label{ss:chp}

The large abundance of interstellar \chp\ has been a puzzle for decades since its discovery in 1941. Steady-state models where ultraviolet irradiation from stars drives the chemistry in the clouds do not nearly produce the observed \chp\ abundances. The ingredients \cp\ and \hh\ are readily available in diffuse clouds, but their direct reaction is highly endothermic so that at $T \sim 100$\,K, the production of \chp\ must proceed via a slow radiative association reaction. Non-thermal formation mechanisms have been proposed, using turbulence as an energy source to overcome the barrier of the reaction of \cp\ with \hh\ \citep{godard2009}. These models could only be tested indirectly so far because from the ground, only optical absorption lines can be observed at limited velocity resolution.

The rotational transitions of \chp\ are inaccessible from the ground, although the $^{13}$\chp\ $J$=1--0 line has been detected from Mauna Kea \citep{falgarone2005}. With HIFI, both the 1--0 and the 2--1 transitions of \chp\ have been observed toward a number of sources \citep{falgarone2010a,bruderer2010}. The strong absorption lines of \chp\ and $^{13}$\chp\ confirm the earlier abundance estimates of \chp/\hh\ = $\sim$\pow{3}{-8} from the optical spectra. Models of so-called turbulent dissipation regions (TDRs) where the dissipation of turbulence through shocks acts as energy source, reproduce these abundances and also the observed \chp/HCO$^+$ ratios toward interstellar clouds \citep{falgarone2010b}.

In the case of high-mass protostellar envelopes, the lines of \chp\ appear in emission, which suggests a different formation mechanism. The models presented by \citet{bruderer2010} explain the observed abundance and excitation of CH and \chp\ toward the source AFGL 2591 with a scenario of irradiated outflow walls, where a cavity etched out by the outflow allows protostellar far-ultraviolet photons to irradiate and heat the envelope to large distances driving the chemical reactions that produce these molecules.

\subsection{Chlorine chemistry}
\label{ss:cl}




The dominant type of chemical reaction in the gas phase at low temperatures ($\ltsim$100\,K) are ion-molecule reactions, which usually have no activation barrier and proceed at the Langevin rate. The chemistry of interstellar chlorine is a good test case of this ion-molecule reaction scheme. The dominant form of chlorine in diffuse interstellar clouds is Cl$^+$, which reacts exothermically with \hh. The product HCl$^+$ reacts again with \hh, and the resulting H$_2$Cl$^+$ produces HCl upon dissociative recombination with an electron. This scheme is similar to the chemistry of carbon, with the difference that all reactions are exothermic. However, of the chlorine hydrides, only HCl is observable from the ground \citep{blake1985,peng2010}. By giving access to HCl$^+$ and H$_2$Cl$^+$, HIFI provides the first test of the above reaction scheme.

Observations with HIFI have resulted in the interstellar detections of both HCl$^+$ \citep{deluca2011} and H$_2$Cl$^+$ \citep{lis2010}. The abundance ratios of HCl, HCl$^+$ and  H$_2$Cl$^+$ are in good agreement with expectations, but the absolute abundances are $\sim$10$\times$ higher than the models predict. The origin of this discrepancy is not understood; further observations may help to clarify the picture \citep{neufeld2012}. The chlorine isotopic ratio $^{35}$Cl/$^{37}$Cl of $\approx$2 measured toward the massive star-forming region W3~A is somewhat below the Solar value of $\approx$3, but the difference is probably not significant \citep{cernicharo2010}.

\subsection{Hydrides as radiation diagnostics}
\label{ss:hydr}


One way for young stars to disperse their natal envelopes is by energetic radiation (ultraviolet and X-rays), but the efficiency of this process is not well known because the envelopes are opaque to such short-wavelength radiation. Chemical signatures may help to determine the amount and nature of hidden energetic radiation, and the abundances of certain hydride molecules appear to be particularly good diagnostics. The small reduced masses of such molecules imply high line frequencies, so that HIFI offers the first opportunity of their observation.

A number of hydride lines have been observed towards the high-mass protostars AFGL 2591 \citep{bruderer2010} and W3 IRS5 \citep{benz2010}. Detected species are CH, \chp, \ohp, \hhop, \hhhop, NH and SH$^+$, while upper limits were obtained for SH and NH$^+$. 
The lines are narrow, which combined with the estimated abundances indicates an origin in the walls of outflow cavities which are being irradiated by protostellar far-ultraviolet radiation. 
The ground state lines of \ohp\ and \hhop\ show pure absorption profiles suggesting an origin in the outer envelopes of the young stars or in foreground clouds, depending on their velocity shift. 
Excited state lines of \chp, \ohp\ and \hhop\ show P~Cygni profiles which indicate an origin in a wind, as also suggested by source-to-source variations in \hho/\hhop\ line ratios \citep{wyrowski2010}.
Some emission lines, in particular those of \hho, OH and \ohp, also exhibit a broad component (\dv$\approx$30\,\kms) which likely originates in shocks \citep{wampfler2011}.
In conclusion, the hydride observations indicate that the interaction of young high-mass stars with their environment is dominated by far-UV radiation and shocks, whereas X-rays do not seem to play a role.

\section{Formation of stars and planets}
\label{s:psf}

The stars that make up the bulk of the mass and the luminosity of galaxies have not always existed and will not remain forever. The formation of stars and planets is a central question of modern astrophysics, and a prerequisite to understand the evolution of galaxies. All three instruments onboard the Herschel telescope have star formation as a high scientific priority. The high spectral resolution of HIFI makes it especially suitable to disentangle the rich forest-like line spectra of star-forming regions and infer physical conditions from the myriad of molecular lines, for example through modeling of line ratios. In addition, only HIFI can spectrally resolve the line profiles so that gas motions can be studied in exquisite detail. Finally, the chemical composition of star-forming regions is a powerful tool to study physical processes which are not directly observable, such as ionization and dissociation by cosmic rays and energetic radiation fields. The next subsections give highlights of what HIFI has been able to achieve in these areas so far.

\subsection{Structure of star-forming regions}
\label{ss:sfr}

Understanding the formation of stars requires accurate descriptions of the distribution of physical conditions in star-forming regions. Observations of molecular rotational emission lines are an excellent tool to constrain these conditions, as the emitted line spectrum is sensitive to basic parameters such as kinetic temperature and volume density. A good example is the \meth\ molecule, which due to its asymmetric structure has many of its lines grouped together in bands. Such bands give access to a large range in energy levels within a narrow frequency range, thereby enabling accurate measurements of gas temperatures.

\begin{figure}[tb]
\centering
\includegraphics[width=\textwidth,angle=0]{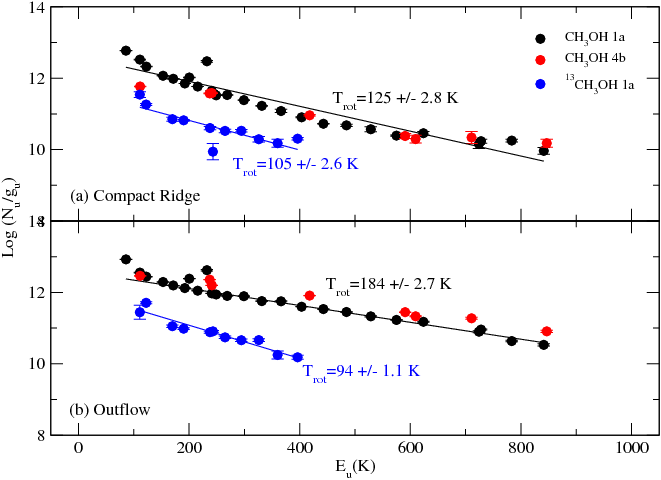}
\caption{Rotation diagram analysis of \meth\ lines toward the Orion-KL region, showing how gas temperatures may be inferred from HIFI multi-line spectra.  From Wang et al (2011).}
\label{f:rd}
\end{figure}

Many of these \meth\ bands, in particular the so-called Q-branches (sets of lines with $\Delta${\it J}=0) are unobservable from the ground, and HIFI is the first opportunity to use their diagnostic value. Examples are the studies of the OMC2-FIR4 core, which is found to be heated from the inside \citep{kama2010}, and of the Compact Ridge in Orion, where external heating appears to dominate \citep{wang2011}; see Figure~\ref{f:rd}.

The capability of HIFI to measure line spectra over a broad wavelength range is also useful to constrain physical conditions in star-forming regions. For example, \citet{yildiz2010} \new{use} high-$J$ lines of CO to measure the temperature profiles of low-mass star-forming cores, and \citet{plume2012} \new{use} C$^{18}$O lines from $J$=5--4 to 17--16 to measure the total column density of CO towards Orion-KL directly in all rotational states, independent of excitation. The populations of the energy levels of C$^{18}$O are found to actually follow thermal distributions, at temperatures consistent with previous estimates for the various components of the Orion-KL region.

\subsection{Shocks and hot cores}
\label{ss:shc}

\begin{figure}[tb]
\centering
\includegraphics[width=0.7\textwidth,angle=0]{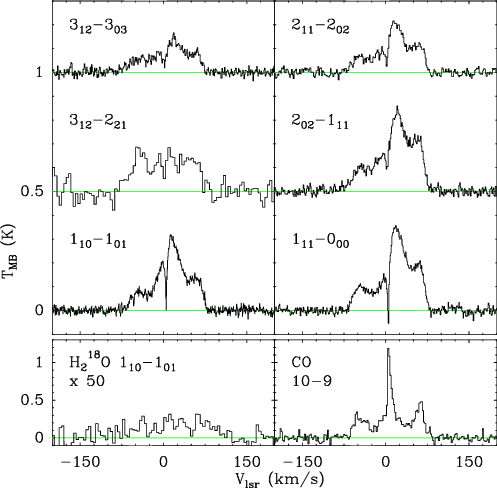}
\caption{HIFI spectra of \hho\ lines toward the low-mass protostar L1448 show high-velocity 'bullets' \new{at \vlsr\ $\approx$--50 and +60\,\kms,} which demonstrate that \hho\ is a good tracer of gas motions. From Kristensen et al (2011).}
\label{f:l1448}
\end{figure}

The \hho\ molecule is a useful tracer of physical conditions in star-forming regions in several ways. Its line ratios can be used as tracers of the kinetic temperature and the volume density of the gas, and its line intensities are a measure of its abundance, which gives information about chemical formation and destruction processes. Towards low-mass star-forming regions, a third method turns out to be particularly powerful: \hho\ line profiles are found to be sensitive tracers of gas dynamics. \citet{kristensen2011} has found `bullets' in HIFI spectra of the outflow of the young protostar L1448 (Figure~\ref{f:l1448}). These high-velocity features in the line profiles were known from ground-based CO observations, but they are much brighter in \hho\ \citep{bjerkeli2011}. Kristensen et al derive high temperatures ($\approx$150\,K) and densities ($\approx$10$^5$\,\ccm) for the outflow shocks where the bullet features are thought to arise. The \hho\ and CO cooling appears to be similar in magnitude, and the high \hho\ abundance indicates that the hydrogen in the shock is mostly in the form of \hh.

\begin{figure}[tb]
\centering
\includegraphics[width=0.7\textwidth,angle=-90]{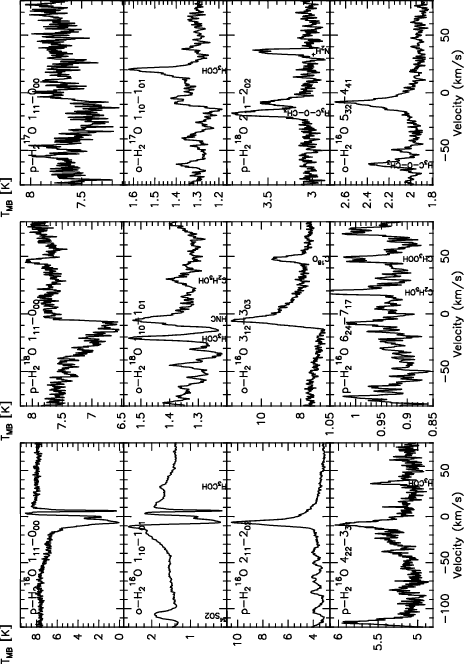}
\caption{Observations with HIFI toward the high-mass protostar NGC 6334I show \hho\ lines from several isotopes and from a wide range of energy levels. The line profiles show a mix of emission and absorption from the protostellar envelope, the outflow, and several foreground clouds. From Emprechtinger et al (2010).}
\label{f:ngc6334}
\end{figure}

Observations of gas-phase molecules toward dense interstellar clouds indicate reduced abundances of all but the lightest species, presumably as a result of depletion onto dust grains.
The bulk material of the grain surfaces is \hho\ ice, as shown by the strong and broad mid-infrared absorption features in the spectra of any cloud with $A_V > 3$ \citep{willner1982,whittet2010}.
When newly formed stars heat up their surroundings, the ice mantles will evaporate and enrich the gas phase with molecules synthesized on the grain surfaces. 
Evidence that this process occurs are observations of complex organic molecules in the spectra of young stars which have heated up their surroundings, the so-called hot cores \citep{herbst2009}. 
Contrary to expectation, however, Herschel-HIFI finds that the \hho\ abundance in hot cores is $\sim$10$^{-6}$ which is only $\sim$1\% of gas-phase oxygen and much less than the \hho\ abundance in the ice mantles. The most accurate measurements are toward high-mass hot cores (Figure~\ref{f:ngc6334}), which are bright enough to see excited state lines and rare isotopic lines of \hho\ \citep{marseille2010,chavarria2010,emprechtinger2010} but the result holds for lower-mass objects as well \citep{vandishoeck2011}. In some cases, observations of high-$J$ lines of rare \hho\ isotopologs reveal higher abundances \citep{melnick2010,herpin2011}, indicating that \new{the low-$J$ lines of \hho\ itself are not sensitive to} the evaporated grain mantle material, but how general this result is remains to be seen.

\subsection{Protoplanetary disks}

\begin{figure}[tb]
\centering
\includegraphics[width=0.7\textwidth,angle=0]{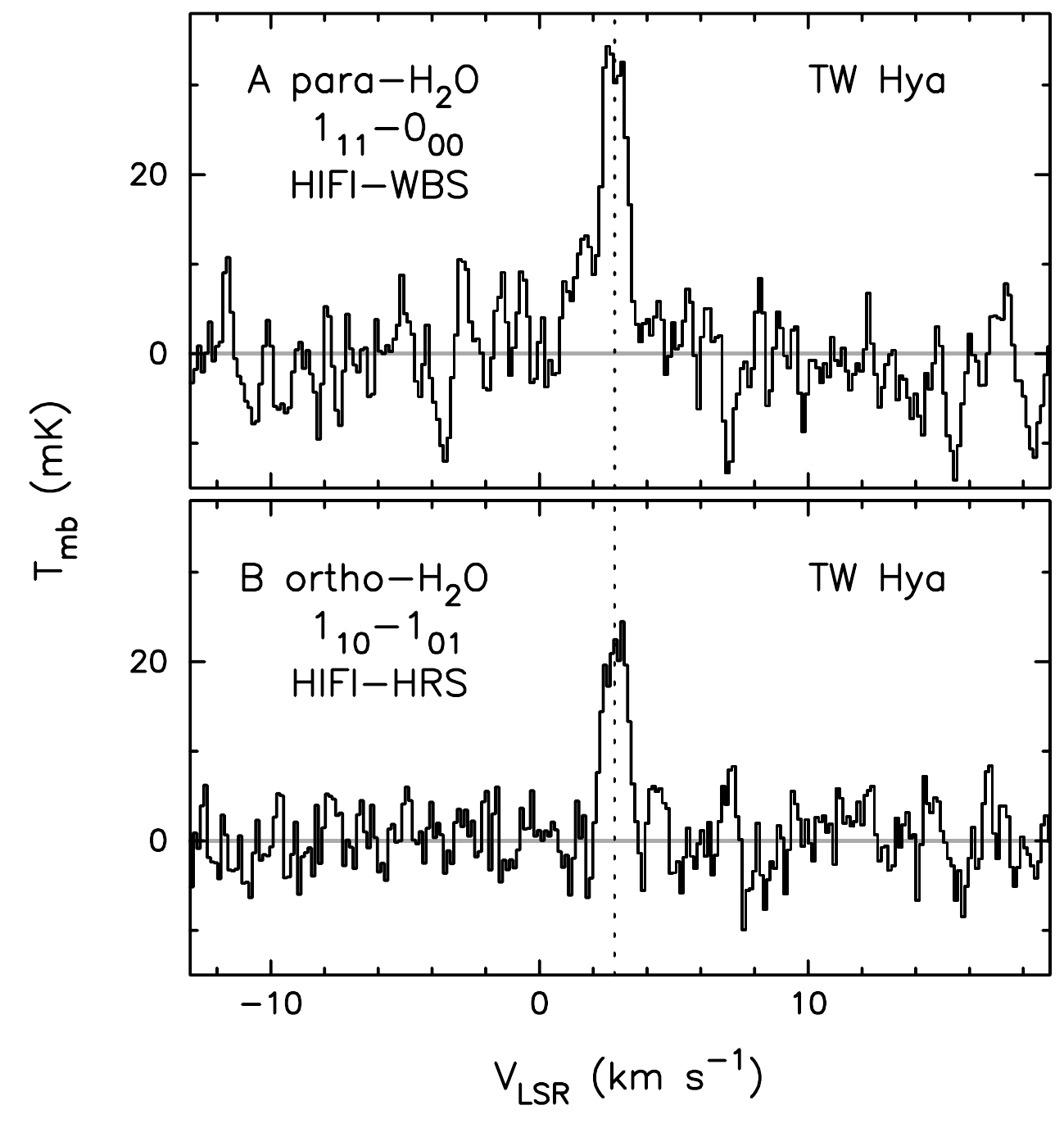}
\caption{Detection of line emission from gas-phase \hho\ in the TW Hya protoplanetary disk with HIFI, which \new{show a low abundance and a low ortho/para ratio of \hho\ } in this forming planetary system. From Hogerheijde et al (2011).}
\label{f:twhya}
\end{figure}

The star formation process leads to a disk of gas and dust which surrounds its parent star for $\sim$10\,Myr after its formation. The structure of such disks is of great interest since they are the likely birthplaces of exoplanets. The gross structure of protoplanetary disks is well known: the density is high in the midplane, where molecules freeze out onto dust grains, and low in the disk atmosphere, where stellar ultraviolet radiation first desorbs and then dissociates the molecules.
The radial structure of the disk is usually described by decreasing (power-law) functions of temperature and density with increasing radius.
However, key parameters of the disks are not well understood, such as the amount of mixing between the various parts of the disk, and the amount of settling of dust grains as a function of time \citep{bergin2007}.

Observations of \hho\ lines with HIFI may improve our understanding of protoplanetary disks, but the expected signals are weak so that only few objects can be observed. The spectrum of DM Tau does not show \hho\ lines at all which gives a firm upper limit on the \hho\ abundance \citep{bergin2010}. The implication of this observation is that the dust in this disk has already settled to the midplane and probably started coagulating, which makes photodesorption less effective. The HIFI data thus imply that planet formation is underway in this object.

Observations of a second disk, TW Hya, have resulted in the detection of both ground-state lines of \hho, as shown in Figure~\ref{f:twhya} \citep{hogerheijde2011}. The lines are likely optically thick, but radiative transfer calculations suggest that not only the abundance of \hho\ in this object is low, but also its ortho/para ratio. In particular, the o/p ratio of \hho\ in the TW Hya disk is lower than that in comets in our Solar System, \new{which is a surprise because comets are thought to originate in the outer protosolar nebula, which is the part that the HIFI data probe. This result suggests that some processing of pre-cometary material occurred in the inner protosolar nebula, for example as a result of mixing of the inner and outer parts of protoplanetary disks.}

\section{Solar system results}
\label{s:solsys}

Planets form in circumstellar disks which are a natural byproduct of star formation. 
As realized by Kant and Laplace in the eighteenth century, the fact that the planets of the Solar system share one orbital plane and revolve around the Sun in the same direction suggests their formation from a protoplanetary disk, and argues against a scenario where the planets were captured one by one.
However, the diversity of objects in our Solar system is large, and no two planets are alike, which implies that the formation of planetary systems is far from a simple process. The origin of this diversity and the \new{likelihood} of habitable planets around other stars are currently major science topics.

Today, most young stars are known to be surrounded by disks \citep{dullemond2010} and a sizeable fraction of mature stars (up to 50\%) are known to host planets \citep{mayor2011}. Studying the disks and planets around other stars remains a challenge, because of the high demands that it sets to instrumentation.
Our Solar system covers only a fraction of parameter space: in particular, hot Jupiters and super-Earths do not occur around the Sun. Nevertheless, studies of the local planets, moons, comets and asteroids give a first impression of the possible outcomes of the planet formation process.  This section describes key results from HIFI concerning physical processes and the chemical composition of Solar system objects.

\subsection{The source of \hho\ to the early Earth}
\label{ss:comet}


\begin{figure}[tb]
\centering
\includegraphics[width=\textwidth,angle=0]{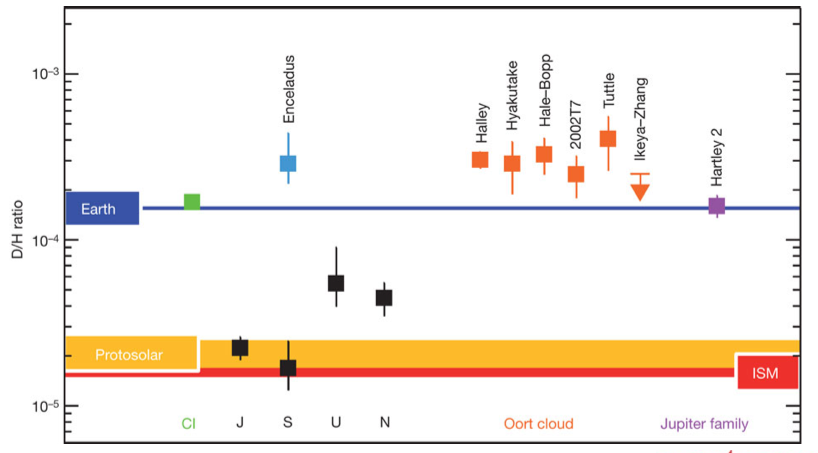}
\caption{The D/H ratio of \hho\ as measured for various bodies in the Solar system. From Hartogh et al (2011).}
\label{f:hartogh}
\end{figure}

A key result of HIFI concerns the D/H ratio of cometary water. The origin of water on Earth is a long-standing question because the early Earth was too warm to retain such volatile species. Delivery by comets during the Great Bombardment is an attractive solution, except that the isotopic composition, in particular the D/H ratio, of water in well-known comets such as Halley, Hale-Bopp and Hyakutake is at least twice the value of \pow{1.56}{-4} for the Earth's oceans (Figure~\ref{f:hartogh}). Therefore, the Nice model for the evolution of the Solar system has most of the volatiles, including \hho, coming from asteroids \citep{morbidelli2010}.
However, the measured comets all stem from the Oort cloud, where they were \new{expelled by gravitational interaction with} the giant planets. 
The D/H ratios of such \new{long-period} comets thus may not reflect pristine conditions but be affected by processing in the inner Solar system. With HIFI, the first measurement has been obtained of the D/H ratio of water in a \new{short-period} comet from the Kuiper belt, which should reflect pristine conditions in the outer Solar system. The measured ratio of \pow{(1.61 $\pm$ 0.24)}{-4} \citep{hartogh2011comet} is consistent with the value for terrestrial ocean water, implying that comets may have delivered at least some of the water to the early Earth. The relative importance of comets and asteroids as sources of volatiles is a subject for future investigation.


\subsection{External supply of \hho\ to the giant planets}
\label{ss:enceladus}

The far-infrared line profiles of \hho\ seen with ISO toward Saturn show broad absorption from tropospheric \hho, as well as narrower emission due to \hho\ in the stratosphere \citep{feuchtgruber1997}. While \hho\ is expected to be the main oxygen carrier in the troposphere, the stratospheric \hho\ must have an external origin. The cryovolcanic activity on Enceladus has been a candidate supplier since its discovery by the Cassini spacecraft, but measurements with its mass spectrometer are inconclusive about the amount of ejected \hho, which is also thought to produce the E~ring \citep{perry2010}.

Observations of \hho\ lines with HIFI have resulted in the first accurate measurement of Enceladus' \hho\ production rate \citep{hartogh2011enceladus}. The torus is found to have an \hho\ column density of \pow{4}{13}\,\scm\ and a scale height of 50,000\,km, which is sufficient to be the major source of \hho\ for Saturn's upper atmosphere, but not for that of Titan. Similarly, HIFI observations of \hho\ toward Jupiter argue for an origin of its stratospheric \hho\ in cometary impacts such as that of Shoemaker-Levy~9 in July 1994 \citep{cavalie2010}.

The ISO satellite also found \hho\ in the atmosphere of Titan, but its low spectral resolution precluded a determination of the vertical profile \citep{coustenis1998}.
Observations with HIFI reveal emission lines of \hho\ and HNC which given the narrow line width must originate in the upper atmosphere, at altitudes above 300\,km \citep{moreno2010}. 
Preliminary analysis suggests that ablation of micrometeorites is the likely source, although cometary impacts and local ring/satellite source may also play a role.
The first detection of HNC on Titan allows to determine the HCN/HNC ratio at high altitudes ($\sim$700\,km),
which helps to constrain photochemical models of Titan's atmosphere \citep{moreno2011}.




\section{The envelopes of evolved stars}
\label{s:agb}

After leaving the main sequence, stars start losing significant amounts of mass and swell up to become red giants of several types. The mass loss reaches its peak during the AGB phase, after which the star traverses the Hertzsprung gap and starts cooling down as a white dwarf. The ejecta form the so-called AGB remnant, which contains up to half the stellar mass. Young white dwarfs have high surface temperatures and ionize their surroundings, which become visible as planetary nebulae. \new{As these nebulae expand, they disappear from view} and the material disperses into the interstellar medium. These late stages of stellar evolution are a key phase in the lifecycle of gas and dust, when enriched material returns to the ISM, from which future generations of stars may form. In addition, the winds of AGB stars have suitable conditions (high density, low temperature) for the formation of dust grains, which play a key role in the physics and chemistry of interstellar matter. The Herschel-HIFI mission has made several key observations of the material surrounding evolved stars, which have improved our understanding of this part of the life cycle of gas and dust.

One crucial observation concerns the origin of \hho\ in circumstellar envelopes. The SWAS mission has discovered gas-phase \hho\ in the envelopes of C-rich AGB stars, which was attributed to the evaporation of comet-like bodies \citep{melnick2001}. The observations of a single spectral line did however not constrain the temperature of the \hho, which was a major uncertainty in this interpretation. Observations with the PACS and HIFI instruments onboard Herschel have now revealed about a dozen \hho\ lines and show that the temperature is at least several 100\,K \citep{decin2010pacs,decin2010hifi}.
The initial observations of the prototype star IRC 10216 are confirmed by a survey of 8 C-rich AGB stars \citep{neufeld2011}. 
The high \hho\ temperature rules out comet evaporation; instead, a possible origin of the observed \hho\ is
the penetration of ultraviolet photons deep into a clumpy circumstellar envelope \citep{decin2010pacs}. This mechanism would also trigger the formation of other molecules such as \ammo, whose observed abundances are much higher than thermal equilibrium models predict \citep{menten2010}. 
Alternatively, non-equilibrium chemistry such as shock processing of the ejecta at the base of the wind, just above the photosphere, predicts \hho\ abundances which are very similar to the observed values \citep{cherchneff2011}.


The HIFI instrument has also been useful to measure the temperatures of the winds in protoplanetary nebulae and young planetary nebulae. \new{\citet{bujarrabal2011}} present observations of CO, \hho, and other species toward ten such objects and find a surprisingly large range of wind temperatures: from $\approx$30\,K to $\gtsim$200\,K. These differences may be due to cooling in shocks and hence reflect the ages of the winds which range from $\sim$100\,yr for the \new{youngest and warmest} winds to $\sim$1000\,yr for the \new{oldest and coolest} ones.


\section{Conclusions and outlook}
\label{s:concl}

The Herschel mission is still ongoing, and current estimates of its lifetime suggest that liquid helium will not run out before February 2013. Nevertheless, it is already possible to draw some first conclusions from the HIFI results obtained so far.
First, it appears that interstellar clouds have a mixed instead of a layered structure. This conclusion follows in particular from the observations of \cp\ which trace `warm dark gas', but also from the ubiquitous nature and high abundances of interstellar \ohp\ and \hhop. In all these cases, the formerly common classification of interstellar clouds into diffuse clouds where hydrogen is largely atomic and dense clouds where it is mostly molecular must be replaced by a new view that the atomic and molecular phases of interstellar clouds are mixed together. The situation is less clear for protoplanetary disks, where the observations of ortho- and para-\hho\ toward TW Hya suggest that comets are the mixing products of the entire disk, while the observed D/H ratio in comet Hartley indicates distinct conditions for the inner and outer parts of the disks. In this area, more observations are needed.

A second major conclusion is that neither \hho\ nor \oo\ is a major carrier of interstellar oxygen, and that both species play only a minor role in the cooling of interstellar clouds. This conclusion strengthens earlier findings from the SWAS and Odin missions. Only in exceptional situations, such as strong shocks, does \hho\ cooling become important.

In the case of Solar system objects, the main conclusion from the HIFI results is that planetary bodies do not lead solitary lives. Interaction plays an important role, as shown by the observations of exchange of \hho\ and other species between comets and asteroids with planets and their moons. The measurement of the HDO/\hho\ ratio in a Jupiter family comet which shows that comets were a major supplier of \hho\ to the early Earth is an example of such interactions in the past of the Solar system.

In the near future, the newly commissioned ALMA interferometer will break new ground on each of these science topics. In particular, the high angular resolution of ALMA will be crucial to understand the structure of small objects such as protoplanetary disks. The good sensitivity of ALMA will be especially important to study distant processes such as star formation in the early Universe.

In the more distant future, several planned telescopes and space missions will build on the legacy of Herschel and HIFI. 
In particular, the CCAT telescope will make deep wide-field images of the submillimeter sky at high ($\sim$10$''$) resolution, both in continuum and spectral lines.
From $\approx$2015 on, the GUSSTO balloon experiment will survey the Southern sky in the fine structure lines of O, \cp\ and N$^+$ at high spectral resolution.
The SAFARI instrument onboard the JAXA-led SPICA mission, to be launched around 2020, will make very deep images of spectral line emission in the far-infrared.
Finally, the FIRI interferometer will bring high angular resolution to the far-infrared regime in the year $\approx$2030.

\section*{Acknowledgements}

The author thanks Frank Helmich, Xander Tielens and Ted Bergin for useful discussions and comments on the manuscript.

HIFI has been designed and built by a consortium of institutes and university departments from across Europe, Canada and the US under the leadership of SRON Netherlands Institute for Space Research, Groningen, The Netherlands with major contributions from Germany, France and the US. Consortium members are: Canada: CSA, U.Waterloo; France: CESR, LAB, LERMA, IRAM; Germany: KOSMA, MPIfR, MPS; Ireland, NUI Maynooth; Italy: ASI, IFSI-INAF, Arcetri-INAF; Netherlands: SRON, TUD; Poland: CAMK, CBK; Spain: Observatorio Astron\'omico Nacional (IGN), Centro de Astrobiolog\'{\i}a (CSIC-INTA); Sweden: Chalmers University of Technology - MC2, RSS \& GARD, Onsala Space Observatory, Swedish National Space Board, Stockholm University - Stockholm Observatory; Switzerland: ETH Z\"urich, FHNW; USA: Caltech, JPL, NHSC.


\bibliographystyle{adspres}
\bibliography{hifi}

\begin{thebibliography}{88}
\expandafter\ifx\csname natexlab\endcsname\relax\def\natexlab#1{#1}\fi

\bibitem[{{Ag{\'u}ndez} {et~al.}(2011){Ag{\'u}ndez}, {Cernicharo}, {Waters},
  {Decin}, {Encrenaz}, {Neufeld}, {Teyssier}, \& {Daniel}}]{agundez2011}
{Ag{\'u}ndez}, M., {Cernicharo}, J., {Waters}, L.~B.~F.~M., {et~al.} 2011,
  \aap, 533, L6

\newblock {HIFI detection of hydrogen fluoride in the carbon star envelope IRC
  +10216}.
\newblock {\em \aap\/}~{\bf 533}, L6.

\bibitem[{{Benz} {et~al.}(2010){Benz}, {Bruderer}, {van Dishoeck},
  {St{\"a}uber}, {Wampfler}, {Melchior}, {Dedes}, {Wyrowski}, {Doty}, {van der
  Tak}, {B{\"a}chtold}, {Csillaghy}, {Megej}, {Monstein}, {Soldati},
  {Bachiller}, {Baudry}, {Benedettini}, {Bergin}, {Bjerkeli}, {Blake},
  {Bontemps}, {Braine}, {Caselli}, {Cernicharo}, {Codella}, {Daniel}, {di
  Giorgio}, {Dieleman}, {Dominik}, {Encrenaz}, {Fich}, {Fuente}, {Giannini},
  {Goicoechea}, {de Graauw}, {Helmich}, {Herczeg}, {Herpin}, {Hogerheijde},
  {Jacq}, {Jellema}, {Johnstone}, {J{\o}rgensen}, {Kristensen}, {Larsson},
  {Lis}, {Liseau}, {Marseille}, {McCoey}, {Melnick}, {Neufeld}, {Nisini},
  {Olberg}, {Ossenkopf}, {Parise}, {Pearson}, {Plume}, {Risacher},
  {Santiago-Garc{\'{\i}}a}, {Saraceno}, {Schieder}, {Shipman}, {Stutzki},
  {Tafalla}, {Tielens}, {van Kempen}, {Visser}, \& {Y{\i}ld{\i}z}}]{benz2010}
{Benz}, A.~O., {Bruderer}, S., {van Dishoeck}, E.~F., {et~al.} 2010, \aap, 521,
  L35

\newblock {Hydrides in young stellar objects: Radiation tracers in a
  protostar-disk-outflow system}.
\newblock {\em \aap\/}~{\bf 521}, L35.

\bibitem[{{Bergin} {et~al.}(2007){Bergin}, {Aikawa}, {Blake}, \& {van
  Dishoeck}}]{bergin2007}
{Bergin}, E.~A., {Aikawa}, Y., {Blake}, G.~A., \& {van Dishoeck}, E.~F. 2007,
  Protostars and Planets V, 751

\newblock {The Chemical Evolution of Protoplanetary Disks}.
\newblock {\em Protostars and Planets V\/}, 751--766.

\bibitem[{{Bergin} {et~al.}(2010){Bergin}, {Hogerheijde}, {Brinch}, {Fogel},
  {Y{\i}ld{\i}z}, {Kristensen}, {van Dishoeck}, {Bell}, {Blake}, {Cernicharo},
  {Dominik}, {Lis}, {Melnick}, {Neufeld}, {Pani{\'c}}, {Pearson}, {Bachiller},
  {Baudry}, {Benedettini}, {Benz}, {Bjerkeli}, {Bontemps}, {Braine},
  {Bruderer}, {Caselli}, {Codella}, {Daniel}, {di Giorgio}, {Doty}, {Encrenaz},
  {Fich}, {Fuente}, {Giannini}, {Goicoechea}, {de Graauw}, {Helmich},
  {Herczeg}, {Herpin}, {Jacq}, {Johnstone}, {J{\o}rgensen}, {Larsson},
  {Liseau}, {Marseille}, {McCoey}, {Nisini}, {Olberg}, {Parise}, {Plume},
  {Risacher}, {Santiago-Garc{\'{\i}}a}, {Saraceno}, {Shipman}, {Tafalla}, {van
  Kempen}, {Visser}, {Wampfler}, {Wyrowski}, {van der Tak}, {Jellema},
  {Tielens}, {Hartogh}, {St{\"u}tzki}, \& {Szczerba}}]{bergin2010}
{Bergin}, E.~A., {Hogerheijde}, M.~R., {Brinch}, C., {et~al.} 2010, \aap, 521,
  L33

\newblock {Sensitive limits on the abundance of cold water vapor in the DM
  Tauri protoplanetary disk}.
\newblock {\em \aap\/}~{\bf 521}, L33.

\bibitem[{{Bergin} {et~al.}(1995){Bergin}, {Langer}, \&
  {Goldsmith}}]{bergin1995}
{Bergin}, E.~A., {Langer}, W.~D., \& {Goldsmith}, P.~F. 1995, \apj, 441, 222

\newblock {Gas-phase chemistry in dense interstellar clouds including grain
  surface molecular depletion and desorption}.
\newblock {\em \apj\/}~{\bf 441}, 222--243.

\bibitem[{{Bergin} {et~al.}(2000){Bergin}, {Melnick}, {Stauffer}, {Ashby},
  {Chin}, {Erickson}, {Goldsmith}, {Harwit}, {Howe}, {Kleiner}, {Koch},
  {Neufeld}, {Patten}, {Plume}, {Schieder}, {Snell}, {Tolls}, {Wang},
  {Winnewisser}, \& {Zhang}}]{bergin2000}
{Bergin}, E.~A., {Melnick}, G.~J., {Stauffer}, J.~R., {et~al.} 2000, \apjl,
  539, L129

\newblock {Implications of Submillimeter Wave Astronomy Satellite Observations
  for Interstellar Chemistry and Star Formation}.
\newblock {\em \apjl\/}~{\bf 539}, L129--L132.

\bibitem[{{Bjerkeli} {et~al.}(2011){Bjerkeli}, {Liseau}, {Nisini}, {Tafalla},
  {Benedettini}, {Bergman}, {Dionatos}, {Giannini}, {Herczeg}, {Justtanont},
  {Larsson}, {McOey}, {Olberg}, \& {Olofsson}}]{bjerkeli2011}
{Bjerkeli}, P., {Liseau}, R., {Nisini}, B., {et~al.} 2011, \aap, 533, A80

\newblock {Herschel observations of the Herbig-Haro objects HH 52-54}.
\newblock {\em \aap\/}~{\bf 533}, A80.

\bibitem[{{Blake} {et~al.}(1985){Blake}, {Keene}, \& {Phillips}}]{blake1985}
{Blake}, G.~A., {Keene}, J., \& {Phillips}, T.~G. 1985, \apj, 295, 501

\newblock {Chlorine in dense interstellar clouds - The abundance of HCl in
  OMC-1}.
\newblock {\em \apj\/}~{\bf 295}, 501--506.

\bibitem[{{Bruderer} {et~al.}(2010){Bruderer}, {Benz}, {van Dishoeck},
  {Melchior}, {Doty}, {van der Tak}, {St{\"a}uber}, {Wampfler}, {Dedes},
  {Y{\i}ld{\i}z}, {Pagani}, {Giannini}, {de Graauw}, {Whyborn}, {Teyssier},
  {Jellema}, {Shipman}, {Schieder}, {Honingh}, {Caux}, {B{\"a}chtold},
  {Csillaghy}, {Monstein}, {Bachiller}, {Baudry}, {Benedettini}, {Bergin},
  {Bjerkeli}, {Blake}, {Bontemps}, {Braine}, {Caselli}, {Cernicharo},
  {Codella}, {Daniel}, {di Giorgio}, {Dominik}, {Encrenaz}, {Fich}, {Fuente},
  {Goicoechea}, {Helmich}, {Herczeg}, {Herpin}, {Hogerheijde}, {Jacq},
  {Johnstone}, {J{\o}rgensen}, {Kristensen}, {Larsson}, {Lis}, {Liseau},
  {Marseille}, {McCoey}, {Melnick}, {Neufeld}, {Nisini}, {Olberg}, {Parise},
  {Pearson}, {Plume}, {Risacher}, {Santiago-Garc{\'{\i}}a}, {Saraceno},
  {Shipman}, {Tafalla}, {van Kempen}, {Visser}, \& {Wyrowski}}]{bruderer2010}
{Bruderer}, S., {Benz}, A.~O., {van Dishoeck}, E.~F., {et~al.} 2010, \aap, 521,
  L44+

\newblock {Herschel/HIFI detections of hydrides towards AFGL 2591. Envelope
  emission versus tenuous cloud absorption}.
\newblock {\em \aap\/}~{\bf 521}, L44+.

\bibitem[{{Bujarrabal} {et~al.}(2012){Bujarrabal}, {Alcolea}, {Soria-Ruiz},
  {Planesas}, {Teyssier}, {Cernicharo}, {Decin}, {Dominik}, {Justtanont}, {de
  Koter}, {Marston}, {Melnick}, {Menten}, {Neufeld}, {Olofsson}, {Schmidt},
  {Sch{\"o}ier}, {Szczerba}, \& {Waters}}]{bujarrabal2011}
{Bujarrabal}, V., {Alcolea}, J., {Soria-Ruiz}, R., {et~al.} 2012, \aap, 537, A8

\newblock {Herschel/HIFI observations of molecular emission in protoplanetary
  nebulae and young planetary nebulae}.
\newblock {\em \aap\/}~{\bf 537}, A8.

\bibitem[{{Cavali{\'e}} {et~al.}(2010){Cavali{\'e}}, {Hartogh}, {Lellouch},
  {Moreno}, {Jarchow}, {Billebaud}, {Orton}, {Rengel}, {Sagawa}, {Lara},
  {Gonzalez}, \& {HssO Team}}]{cavalie2010}
{Cavali{\'e}}, T., {Hartogh}, P., {Lellouch}, E., {et~al.} 2010, in Bulletin of
  the American Astronomical Society, Vol.~42, AAS/Division for Planetary
  Sciences Meeting Abstracts \#42, 1010

\newblock {Mapping water in Jupiter with Herschel/HIFI}.
\newblock In {\em AAS/Division for Planetary Sciences Meeting Abstracts \#42},
  Volume~42 of {\em Bulletin of the American Astronomical Society}, pp.\  1010.

\bibitem[{{Cernicharo} {et~al.}(2010){Cernicharo}, {Goicoechea}, {Daniel},
  {Ag{\'u}ndez}, {Caux}, {de Graauw}, {de Jonge}, {Kester}, {Leduc},
  {Steinmetz}, {Stutzki}, \& {Ward}}]{cernicharo2010}
{Cernicharo}, J., {Goicoechea}, J.~R., {Daniel}, F., {et~al.} 2010, \aap, 518,
  L115

\newblock {The $^{35}$Cl/$^{37}$Cl isotopic ratio in dense molecular clouds:
  HIFI observations of hydrogen chloride towards W3 A}.
\newblock {\em \aap\/}~{\bf 518}, L115.

\bibitem[{{Chavarr{\'{\i}}a} {et~al.}(2010){Chavarr{\'{\i}}a}, {Herpin},
  {Jacq}, {Braine}, {Bontemps}, {Baudry}, {Marseille}, {van der Tak},
  {Pietropaoli}, {Wyrowski}, {Shipman}, {Frieswijk}, {van Dishoeck},
  {Cernicharo}, {Bachiller}, {Benedettini}, {Benz}, {Bergin}, {Bjerkeli},
  {Blake}, {Bruderer}, {Caselli}, {Codella}, {Daniel}, {di Giorgio}, {Dominik},
  {Doty}, {Encrenaz}, {Fich}, {Fuente}, {Giannini}, {Goicoechea}, {de Graauw},
  {Hartogh}, {Helmich}, {Herczeg}, {Hogerheijde}, {Johnstone}, {J{\o}rgensen},
  {Kristensen}, {Larsson}, {Lis}, {Liseau}, {McCoey}, {Melnick}, {Nisini},
  {Olberg}, {Parise}, {Pearson}, {Plume}, {Risacher}, {Santiago-Garc{\'{\i}}a},
  {Saraceno}, {Stutzki}, {Szczerba}, {Tafalla}, {Tielens}, {van Kempen},
  {Visser}, {Wampfler}, {Willem}, \& {Y{\i}ld{\i}z}}]{chavarria2010}
{Chavarr{\'{\i}}a}, L., {Herpin}, F., {Jacq}, T., {et~al.} 2010, \aap, 521, L37

\newblock {Water in massive star-forming regions: HIFI observations of W3
  IRS5}.
\newblock {\em \aap\/}~{\bf 521}, L37.

\bibitem[{{Cherchneff}(2011)}]{cherchneff2011}
{Cherchneff}, I. 2011, \aap, 526, L11

\newblock {Water in IRC+10216: a genuine formation process by shock-induced
  chemistry in the inner wind}.
\newblock {\em \aap\/}~{\bf 526}, L11.

\bibitem[{{Coustenis} {et~al.}(1998){Coustenis}, {Salama}, {Lellouch},
  {Encrenaz}, {Bjoraker}, {Samuelson}, {de Graauw}, {Feuchtgruber}, \&
  {Kessler}}]{coustenis1998}
{Coustenis}, A., {Salama}, A., {Lellouch}, E., {et~al.} 1998, \aap, 336, L85

\newblock {Evidence for water vapor in Titan's atmosphere from ISO/SWS data}.
\newblock {\em \aap\/}~{\bf 336}, L85--L89.

\bibitem[{{De Graauw} {et~al.}(2010){De Graauw}, {Helmich}, {Phillips},
  {Stutzki}, {Caux}, {Whyborn}, {Dieleman}, {Roelfsema}, {Aarts}, {Assendorp},
  {Bachiller}, {Baechtold}, {Barcia}, {Beintema}, {Belitsky}, {Benz}, {Bieber},
  {Boogert}, {Borys}, {Bumble}, {Ca{\"i}s}, {Caris}, {Cerulli-Irelli},
  {Chattopadhyay}, {Cherednichenko}, {Ciechanowicz}, {Coeur-Joly}, {Comito},
  {Cros}, {de Jonge}, {de Lange}, {Delforges}, {Delorme}, {den Boggende},
  {Desbat}, {Diez-Gonz{\'a}lez}, {di Giorgio}, {Dubbeldam}, {Edwards},
  {Eggens}, {Erickson}, {Evers}, {Fich}, {Finn}, {Franke}, {Gaier}, {Gal},
  {Gao}, {Gallego}, {Gauffre}, {Gill}, {Glenz}, {Golstein}, {Goulooze},
  {Gunsing}, {G{\"u}sten}, {Hartogh}, {Hatch}, {Higgins}, {Honingh}, {Huisman},
  {Jackson}, {Jacobs}, {Jacobs}, {Jarchow}, {Javadi}, {Jellema}, {Justen},
  {Karpov}, {Kasemann}, {Kawamura}, {Keizer}, {Kester}, {Klapwijk}, {Klein},
  {Kollberg}, {Kooi}, {Kooiman}, {Kopf}, {Krause}, {Krieg}, {Kramer},
  {Kruizenga}, {Kuhn}, {Laauwen}, {Lai}, {Larsson}, {Leduc}, {Leinz}, {Lin},
  {Liseau}, {Liu}, {Loose}, {L{\'o}pez-Fernandez}, {Lord}, {Luinge}, {Marston},
  {Mart{\'{\i}}n-Pintado}, {Maestrini}, {Maiwald}, {McCoey}, {Mehdi}, {Megej},
  {Melchior}, {Meinsma}, {Merkel}, {Michalska}, {Monstein}, {Moratschke},
  {Morris}, {Muller}, {Murphy}, {Naber}, {Natale}, {Nowosielski}, {Nuzzolo},
  {Olberg}, {Olbrich}, {Orfei}, {Orleanski}, {Ossenkopf}, {Peacock}, {Pearson},
  {Peron}, {Phillip-May}, {Piazzo}, {Planesas}, {Rataj}, {Ravera}, {Risacher},
  {Salez}, {Samoska}, {Saraceno}, {Schieder}, {Schlecht}, {Schl{\"o}der},
  {Schm{\"u}lling}, {Schultz}, {Schuster}, {Siebertz}, {Smit}, {Szczerba},
  {Shipman}, {Steinmetz}, {Stern}, {Stokroos}, {Teipen}, {Teyssier}, {Tils},
  {Trappe}, {van Baaren}, {van Leeuwen}, {van de Stadt}, {Visser}, {Wildeman},
  {Wafelbakker}, {Ward}, {Wesselius}, {Wild}, {Wulff}, {Wunsch}, {Tielens},
  {Zaal}, {Zirath}, {Zmuidzinas}, \& {Zwart}}]{degraauw2010}
{De Graauw}, T., {Helmich}, F.~P., {Phillips}, T.~G., {et~al.} 2010, \aap, 518,
  L6

\newblock {The Herschel-Heterodyne Instrument for the Far-Infrared (HIFI)}.
\newblock {\em \aap\/}~{\bf 518}, L6.

\bibitem[{{De Luca} {et~al.}(2011){De Luca}, {Gupta}, {Neufeld}, {Gerin},
  {Teyssier}, {Pearson}, {Lis}, \& {Herschel Prismas Team}}]{deluca2011}
{De Luca}, M., {Gupta}, H., {Neufeld}, D., {et~al.} 2011, in IAU Symposium,
  Vol. 280, IAU Symposium, 147P

\newblock {Tentative detection of HCl$^+$ in diffuse clouds}.
\newblock In {\em IAU Symposium}, Volume 280 of {\em IAU Symposium}, pp.\
  147P.

\bibitem[{{De Pater} \& {Lissauer}(2001)}]{depater-lissauer}
{De Pater}, I. \& {Lissauer}, J.~J. 2001, {Planetary Sciences} (Cambridge
  University Press)

\newblock {\em {Planetary Sciences}}.
\newblock Cambridge University Press.

\bibitem[{{Decin} {et~al.}(2010{\natexlab{a}}){Decin}, {Ag{\'u}ndez}, {Barlow},
  {Daniel}, {Cernicharo}, {Lombaert}, {De Beck}, {Royer}, {Vandenbussche},
  {Wesson}, {Polehampton}, {Blommaert}, {De Meester}, {Exter}, {Feuchtgruber},
  {Gear}, {Gomez}, {Groenewegen}, {Gu{\'e}lin}, {Hargrave}, {Huygen}, {Imhof},
  {Ivison}, {Jean}, {Kahane}, {Kerschbaum}, {Leeks}, {Lim}, {Matsuura},
  {Olofsson}, {Posch}, {Regibo}, {Savini}, {Sibthorpe}, {Swinyard}, {Yates}, \&
  {Waelkens}}]{decin2010pacs}
{Decin}, L., {Ag{\'u}ndez}, M., {Barlow}, M.~J., {et~al.} 2010{\natexlab{a}},
  \nat, 467, 64

\newblock {Warm water vapour in the sooty outflow from a luminous carbon star}.
\newblock {\em \nat\/}~{\bf 467}, 64--67.

\bibitem[{{Decin} {et~al.}(2010{\natexlab{b}}){Decin}, {Justtanont}, {De Beck},
  {Lombaert}, {de Koter}, {Waters}, {Marston}, {Teyssier}, {Sch{\"o}ier},
  {Bujarrabal}, {Alcolea}, {Cernicharo}, {Dominik}, {Melnick}, {Menten},
  {Neufeld}, {Olofsson}, {Planesas}, {Schmidt}, {Szczerba}, {de Graauw},
  {Helmich}, {Roelfsema}, {Dieleman}, {Morris}, {Gallego},
  {D{\'{\i}}ez-Gonz{\'a}lez}, \& {Caux}}]{decin2010hifi}
{Decin}, L., {Justtanont}, K., {De Beck}, E., {et~al.} 2010{\natexlab{b}},
  \aap, 521, L4

\newblock {Water content and wind acceleration in the envelope around the
  oxygen-rich AGB star IK Tauri as seen by Herschel/HIFI}.
\newblock {\em \aap\/}~{\bf 521}, L4.

\bibitem[{{Draine}(2011)}]{draine2011}
{Draine}, B.~T. 2011, {Physics of the Interstellar and Intergalactic Medium}
  (Princeton University Press)

\newblock {\em {Physics of the Interstellar and Intergalactic Medium}}.
\newblock Princeton University Press.

\bibitem[{{Dullemond} \& {Monnier}(2010)}]{dullemond2010}
{Dullemond}, C.~P. \& {Monnier}, J.~D. 2010, \araa, 48, 205

\newblock {The Inner Regions of Protoplanetary Disks}.
\newblock {\em \araa\/}~{\bf 48}, 205--239.

\bibitem[{{Emprechtinger} {et~al.}(2010){Emprechtinger}, {Lis}, {Bell},
  {Phillips}, {Schilke}, {Comito}, {Rolffs}, {van der Tak}, {Ceccarelli},
  {Aarts}, {Bacmann}, {Baudry}, {Benedettini}, {Bergin}, {Blake}, {Boogert},
  {Bottinelli}, {Cabrit}, {Caselli}, {Castets}, {Caux}, {Cernicharo},
  {Codella}, {Coutens}, {Crimier}, {Demyk}, {Dominik}, {Encrenaz}, {Falgarone},
  {Fuente}, {Gerin}, {Goldsmith}, {Helmich}, {Hennebelle}, {Henning}, {Herbst},
  {Hily-Blant}, {Jacq}, {Kahane}, {Kama}, {Klotz}, {Kooi}, {Langer}, {Lefloch},
  {Loose}, {Lord}, {Lorenzani}, {Maret}, {Melnick}, {Neufeld}, {Nisini},
  {Ossenkopf}, {Pacheco}, {Pagani}, {Parise}, {Pearson}, {Risacher}, {Salez},
  {Saraceno}, {Schuster}, {Stutzki}, {Tielens}, {van der Wiel}, {Vastel},
  {Viti}, {Wakelam}, {Walters}, {Wyrowski}, \& {Yorke}}]{emprechtinger2010}
{Emprechtinger}, M., {Lis}, D.~C., {Bell}, T., {et~al.} 2010, \aap, 521, L28

\newblock {The distribution of water in the high-mass star-forming region NGC
  6334 I}.
\newblock {\em \aap\/}~{\bf 521}, L28.

\bibitem[{{Falgarone} {et~al.}(2010{\natexlab{a}}){Falgarone}, {Godard},
  {Cernicharo}, {de Luca}, {Gerin}, {Phillips}, {Black}, {Lis}, {Bell},
  {Boulanger}, {Coutens}, {Dartois}, {Encrenaz}, {Giesen}, {Goicoechea},
  {Goldsmith}, {Gupta}, {Gry}, {Hennebelle}, {Herbst}, {Hily-Blant}, {Joblin},
  {Ka{\'z}mierczak}, {Ko{\l}os}, {Kre{\l}owski}, {Martin-Pintado}, {Monje},
  {Mookerjea}, {Neufeld}, {Perault}, {Pearson}, {Persson}, {Plume}, {Salez},
  {Schmidt}, {Sonnentrucker}, {Stutzki}, {Teyssier}, {Vastel}, {Yu}, {Menten},
  {Geballe}, {Schlemmer}, {Shipman}, {Tielens}, {Philipp}, {Cros},
  {Zmuidzinas}, {Samoska}, {Klein}, {Lorenzani}, {Szczerba}, {P{\'e}ron},
  {Cais}, {Gaufre}, {Cros}, {Ravera}, {Morris}, {Lord}, \&
  {Planesas}}]{falgarone2010b}
{Falgarone}, E., {Godard}, B., {Cernicharo}, J., {et~al.} 2010{\natexlab{a}},
  \aap, 521, L15

\newblock {CH$^{+}$(1-0) and $^{13}$CH$^{+}$(1-0) absorption lines in the
  direction of massive star-forming regions}.
\newblock {\em \aap\/}~{\bf 521}, L15.

\bibitem[{{Falgarone} {et~al.}(2010{\natexlab{b}}){Falgarone}, {Ossenkopf},
  {Gerin}, {Lesaffre}, {Godard}, {Pearson}, {Cabrit}, {Joblin}, {Benz},
  {Boulanger}, {Fuente}, {G{\"u}sten}, {Harris}, {Klein}, {Kramer}, {Lord},
  {Martin}, {Martin-Pintado}, {Neufeld}, {Phillips}, {R{\"o}llig}, {Simon},
  {Stutzki}, {van der Tak}, {Teyssier}, {Yorke}, {Erickson}, {Fich}, {Jellema},
  {Marston}, {Risacher}, {Salez}, \& {Schm{\"u}lling}}]{falgarone2010a}
{Falgarone}, E., {Ossenkopf}, V., {Gerin}, M., {et~al.} 2010{\natexlab{b}},
  \aap, 518, L118

\newblock {Strong CH$^{+}$ J = 1-0 emission and absorption in DR21}.
\newblock {\em \aap\/}~{\bf 518}, L118.

\bibitem[{{Falgarone} {et~al.}(2005){Falgarone}, {Phillips}, \&
  {Pearson}}]{falgarone2005}
{Falgarone}, E., {Phillips}, T.~G., \& {Pearson}, J.~C. 2005, \apjl, 634, L149

\newblock {First Detection of $^{13}$CH$^{+}$ (J=1-0)}.
\newblock {\em \apjl\/}~{\bf 634}, L149--L152.

\bibitem[{{Feuchtgruber} {et~al.}(1997){Feuchtgruber}, {Lellouch}, {de Graauw},
  {B{\'e}zard}, {Encrenaz}, \& {Griffin}}]{feuchtgruber1997}
{Feuchtgruber}, H., {Lellouch}, E., {de Graauw}, T., {et~al.} 1997, \nat, 389,
  159

\newblock {External supply of oxygen to the atmospheres of the giant planets}.
\newblock {\em \nat\/}~{\bf 389}, 159--162.

\bibitem[{{Gerin} {et~al.}(2010){Gerin}, {de Luca}, {Black}, {Goicoechea},
  {Herbst}, {Neufeld}, {Falgarone}, {Godard}, {Pearson}, {Lis}, {Phillips},
  {Bell}, {Sonnentrucker}, {Boulanger}, {Cernicharo}, {Coutens}, {Dartois},
  {Encrenaz}, {Giesen}, {Goldsmith}, {Gupta}, {Gry}, {Hennebelle},
  {Hily-Blant}, {Joblin}, {Kazmierczak}, {Kolos}, {Krelowski},
  {Martin-Pintado}, {Monje}, {Mookerjea}, {Perault}, {Persson}, {Plume},
  {Rimmer}, {Salez}, {Schmidt}, {Stutzki}, {Teyssier}, {Vastel}, {Yu},
  {Contursi}, {Menten}, {Geballe}, {Schlemmer}, {Shipman}, {Tielens},
  {Philipp-May}, {Cros}, {Zmuidzinas}, {Samoska}, {Klein}, \&
  {Lorenzani}}]{gerin2010}
{Gerin}, M., {de Luca}, M., {Black}, J., {et~al.} 2010, \aap, 518, L110

\newblock {Interstellar OH$^{+}$, H$_{2}$O$^{+}$ and H$_{3}$O$^{+}$ along the
  sight-line to G10.6-0.4}.
\newblock {\em \aap\/}~{\bf 518}, L110.

\bibitem[{{Godard} {et~al.}(2009){Godard}, {Falgarone}, \& {Pineau des
  For{\^e}ts}}]{godard2009}
{Godard}, B., {Falgarone}, E., \& {Pineau des For{\^e}ts}, G. 2009, \aap, 495,
  847

\newblock {Models of turbulent dissipation regions in the diffuse interstellar
  medium}.
\newblock {\em \aap\/}~{\bf 495}, 847--867.

\bibitem[{{Goldsmith} \& {Langer}(1978)}]{goldsmith1978}
{Goldsmith}, P.~F. \& {Langer}, W.~D. 1978, \apj, 222, 881

\newblock {Molecular cooling and thermal balance of dense interstellar clouds}.
\newblock {\em \apj\/}~{\bf 222}, 881--895.

\bibitem[{{Goldsmith} {et~al.}(2011){Goldsmith}, {Liseau}, {Bell}, {Black},
  {Chen}, {Hollenbach}, {Kaufman}, {Li}, {Lis}, {Melnick}, {Neufeld}, {Pagani},
  {Snell}, {Benz}, {Bergin}, {Bruderer}, {Caselli}, {Caux}, {Encrenaz},
  {Falgarone}, {Gerin}, {Goicoechea}, {Hjalmarson}, {Larsson}, {Le Bourlot},
  {Le Petit}, {De Luca}, {Nagy}, {Roueff}, {Sandqvist}, {van der Tak}, {van
  Dishoeck}, {Vastel}, {Viti}, \& {Y{\i}ld{\i}z}}]{goldsmith2011}
{Goldsmith}, P.~F., {Liseau}, R., {Bell}, T.~A., {et~al.} 2011, \apj, 737, 96

\newblock {Herschel Measurements of Molecular Oxygen in Orion}.
\newblock {\em \apj\/}~{\bf 737}, 96.

\bibitem[{{Gonz{\'a}lez-Alfonso} {et~al.}(2008){Gonz{\'a}lez-Alfonso}, {Smith},
  {Ashby}, {Fischer}, {Spinoglio}, \& {Grundy}}]{gonzalez2008}
{Gonz{\'a}lez-Alfonso}, E., {Smith}, H.~A., {Ashby}, M.~L.~N., {et~al.} 2008,
  \apj, 675, 303

\newblock {High-excitation OH and H$_{2}$O Lines in Markarian 231: The
  Molecular Signatures of Compact Far-infrared Continuum Sources}.
\newblock {\em \apj\/}~{\bf 675}, 303--315.

\bibitem[{{Grenier} {et~al.}(2005){Grenier}, {Casandjian}, \&
  {Terrier}}]{grenier2005}
{Grenier}, I.~A., {Casandjian}, J.-M., \& {Terrier}, R. 2005, Science, 307,
  1292

\newblock {Unveiling Extensive Clouds of Dark Gas in the Solar Neighborhood}.
\newblock {\em Science\/}~{\bf 307}, 1292--1295.

\bibitem[{{Habing}(1996)}]{habing1996}
{Habing}, H.~J. 1996, \aapr, 7, 97

\newblock {Circumstellar envelopes and Asymptotic Giant Branch stars}.
\newblock {\em \aapr\/}~{\bf 7}, 97--207.

\bibitem[{{Hartogh} {et~al.}(2011{\natexlab{a}}){Hartogh}, {Lellouch},
  {Moreno}, {Bockel{\'e}e-Morvan}, {Biver}, {Cassidy}, {Rengel}, {Jarchow},
  {Cavali{\'e}}, {Crovisier}, {Helmich}, \& {Kidger}}]{hartogh2011enceladus}
{Hartogh}, P., {Lellouch}, E., {Moreno}, R., {et~al.} 2011{\natexlab{a}}, \aap,
  532, L2

\newblock {Direct detection of the Enceladus water torus with Herschel}.
\newblock {\em \aap\/}~{\bf 532}, L2.

\bibitem[{{Hartogh} {et~al.}(2011{\natexlab{b}}){Hartogh}, {Lis},
  {Bockel{\'e}e-Morvan}, {de Val-Borro}, {Biver}, {K{\"u}ppers},
  {Emprechtinger}, {Bergin}, {Crovisier}, {Rengel}, {Moreno}, {Szutowicz}, \&
  {Blake}}]{hartogh2011comet}
{Hartogh}, P., {Lis}, D.~C., {Bockel{\'e}e-Morvan}, D., {et~al.}
  2011{\natexlab{b}}, \nat, 478, 218

\newblock {Ocean-like water in the Jupiter-family comet 103P/Hartley 2}.
\newblock {\em \nat\/}~{\bf 478}, 218--220.

\bibitem[{{Herbst} \& {van Dishoeck}(2009)}]{herbst2009}
{Herbst}, E. \& {van Dishoeck}, E.~F. 2009, \araa, 47, 427

\newblock {Complex Organic Interstellar Molecules}.
\newblock {\em \araa\/}~{\bf 47}, 427--480.

\bibitem[{Herpin {et~al.}(2011)Herpin, Chavarr\'{\i}a, van~der Tak, 3, 4, 5, 6,
  7, \& 8}]{herpin2011}
Herpin, F., Chavarr\'{\i}a, L., van~der Tak, F., {et~al.} 2011, \aap, submitted

\newblock {The massive protostar W43-MM1 as seen by Herschel-HIFI water
  spectra: high turbulence and accretion luminosity}.
\newblock {\em \aap\/}~{\bf submitted}.

\bibitem[{{Herzberg} \& {Lew}(1974)}]{herzberg1974}
{Herzberg}, G. \& {Lew}, H. 1974, \aap, 31, 123

\newblock {Tentative Identffication of the \hhop\ Ion in Comet Kohoutek.}
\newblock {\em \aap\/}~{\bf 31}, 123.

\bibitem[{{Hogerheijde} {et~al.}(2011){Hogerheijde}, {Bergin}, {Brinch},
  {Cleeves}, {Fogel}, {Blake}, {Dominik}, {Lis}, {Melnick}, {Neufeld},
  {Pani{\'c}}, {Pearson}, {Kristensen}, {Y{\i}ld{\i}z}, \& {van
  Dishoeck}}]{hogerheijde2011}
{Hogerheijde}, M.~R., {Bergin}, E.~A., {Brinch}, C., {et~al.} 2011, Science,
  334, 338

\newblock {Detection of the Water Reservoir in a Forming Planetary System}.
\newblock {\em Science\/}~{\bf 334}, 338.

\bibitem[{{Kama} {et~al.}(2010){Kama}, {Dominik}, {Maret}, {van der Tak},
  {Caux}, {Ceccarelli}, {Fuente}, {Crimier}, {Lord}, {Bacmann}, {Baudry},
  {Bell}, {Benedettini}, {Bergin}, {Blake}, {Boogert}, {Bottinelli}, {Cabrit},
  {Caselli}, {Castets}, {Cernicharo}, {Codella}, {Comito}, {Coutens}, {Demyk},
  {Encrenaz}, {Falgarone}, {Gerin}, {Goldsmith}, {Helmich}, {Hennebelle},
  {Henning}, {Herbst}, {Hily-Blant}, {Jacq}, {Kahane}, {Klotz}, {Langer},
  {Lefloch}, {Lis}, {Lorenzani}, {Melnick}, {Nisini}, {Pacheco}, {Pagani},
  {Parise}, {Pearson}, {Phillips}, {Salez}, {Saraceno}, {Schilke}, {Schuster},
  {Tielens}, {van der Wiel}, {Vastel}, {Viti}, {Wakelam}, {Walters},
  {Wyrowski}, {Yorke}, {Cais}, {G{\"u}sten}, {Philipp}, {Klein}, \&
  {Helmich}}]{kama2010}
{Kama}, M., {Dominik}, C., {Maret}, S., {et~al.} 2010, \aap, 521, L39

\newblock {The methanol lines and hot core of OMC2-FIR4, an intermediate-mass
  protostar, with Herschel/HIFI}.
\newblock {\em \aap\/}~{\bf 521}, L39.

\bibitem[{{Kristensen} {et~al.}(2011){Kristensen}, {van Dishoeck}, {Tafalla},
  {Bachiller}, {Nisini}, {Liseau}, \& {Y{\i}ld{\i}z}}]{kristensen2011}
{Kristensen}, L.~E., {van Dishoeck}, E.~F., {Tafalla}, M., {et~al.} 2011, \aap,
  531, L1

\newblock {Water in low-mass star-forming regions with Herschel (WISH-LM).
  High-velocity H$_{2}$O bullets in L1448-MM observed with HIFI}.
\newblock {\em \aap\/}~{\bf 531}, L1.

\bibitem[{{Langer} {et~al.}(2010){Langer}, {Velusamy}, {Pineda}, {Goldsmith},
  {Li}, \& {Yorke}}]{langer2010}
{Langer}, W.~D., {Velusamy}, T., {Pineda}, J.~L., {et~al.} 2010, \aap, 521, L17

\newblock {C$^{+}$ detection of warm dark gas in diffuse clouds}.
\newblock {\em \aap\/}~{\bf 521}, L17.

\bibitem[{{Larsson} {et~al.}(2007){Larsson}, {Liseau}, {Pagani}, {Bergman},
  {Bernath}, {Biver}, {Black}, {Booth}, {Buat}, {Crovisier}, {Curry},
  {Dahlgren}, {Encrenaz}, {Falgarone}, {Feldman}, {Fich}, {Flor{\'e}n},
  {Fredrixon}, {Frisk}, {Gahm}, {Gerin}, {Hagstr{\"o}m}, {Harju}, {Hasegawa},
  {Hjalmarson}, {Johansson}, {Justtanont}, {Klotz}, {Kyr{\"o}l{\"a}}, {Kwok},
  {Lecacheux}, {Liljestr{\"o}m}, {Llewellyn}, {Lundin}, {M{\'e}gie},
  {Mitchell}, {Murtagh}, {Nordh}, {Nyman}, {Olberg}, {Olofsson}, {Olofsson},
  {Olofsson}, {Persson}, {Plume}, {Rickman}, {Ristorcelli}, {Rydbeck},
  {Sandqvist}, {Sch{\'e}ele}, {Serra}, {Torchinsky}, {Tothill}, {Volk},
  {Wiklind}, {Wilson}, {Winnberg}, \& {Witt}}]{larsson2007}
{Larsson}, B., {Liseau}, R., {Pagani}, L., {et~al.} 2007, \aap, 466, 999

\newblock {Molecular oxygen in the {$\rho$} Ophiuchi cloud}.
\newblock {\em \aap\/}~{\bf 466}, 999--1003.

\bibitem[{{Lis} {et~al.}(2010){Lis}, {Pearson}, {Neufeld}, {Schilke},
  {M{\"u}ller}, {Gupta}, {Bell}, {Comito}, {Phillips}, {Bergin}, {Ceccarelli},
  {Goldsmith}, {Blake}, {Bacmann}, {Baudry}, {Benedettini}, {Benz}, {Black},
  {Boogert}, {Bottinelli}, {Cabrit}, {Caselli}, {Castets}, {Caux},
  {Cernicharo}, {Codella}, {Coutens}, {Crimier}, {Crockett}, {Daniel}, {Demyk},
  {Dominic}, {Dubernet}, {Emprechtinger}, {Encrenaz}, {Falgarone}, {Fuente},
  {Gerin}, {Giesen}, {Goicoechea}, {Helmich}, {Hennebelle}, {Henning},
  {Herbst}, {Hily-Blant}, {Hjalmarson}, {Hollenbach}, {Jack}, {Joblin},
  {Johnstone}, {Kahane}, {Kama}, {Kaufman}, {Klotz}, {Langer}, {Larsson}, {Le
  Bourlot}, {Lefloch}, {Le Petit}, {Li}, {Liseau}, {Lord}, {Lorenzani},
  {Maret}, {Martin}, {Melnick}, {Menten}, {Morris}, {Murphy}, {Nagy}, {Nisini},
  {Ossenkopf}, {Pacheco}, {Pagani}, {Parise}, {P{\'e}rault}, {Plume}, {Qin},
  {Roueff}, {Salez}, {Sandqvist}, {Saraceno}, {Schlemmer}, {Schuster}, {Snell},
  {Stutzki}, {Tielens}, {Trappe}, {van der Tak}, {van der Wiel}, {van
  Dishoeck}, {Vastel}, {Viti}, {Wakelam}, {Walters}, {Wang}, {Wyrowski},
  {Yorke}, {Yu}, {Zmuidzinas}, {Delorme}, {Desbat}, {G{\"u}sten}, {Krieg}, \&
  {Delforge}}]{lis2010}
{Lis}, D.~C., {Pearson}, J.~C., {Neufeld}, D.~A., {et~al.} 2010, \aap, 521, L9

\newblock {Herschel/HIFI discovery of interstellar chloronium
  (H$_{2}$Cl$^{+}$)}.
\newblock {\em \aap\/}~{\bf 521}, L9.

\bibitem[{Liseau {et~al.}(2011)Liseau, Goldsmith, Larsson, 3, 4, 5, 6, 7, \&
  8}]{liseau2011}
Liseau, R., Goldsmith, P., Larsson, B., {et~al.} 2011, \aap, in press,
  arxiv:1202.5637

\newblock {Multi-line detection of O$_2$ toward $\rho$ Oph A}.
\newblock {\em \aap\/}~{\bf in press, arxiv:1202.5637}.

\bibitem[{{Madden} {et~al.}(2011){Madden}, {Galametz}, {Cormier},
  {Lebouteiller}, {Galliano}, {Hony}, {R{\'e}my}, {Sauvage}, {Contursi},
  {Sturm}, {Poglitsch}, {Pohlen}, {Smith}, {Bendo}, \&
  {O'Halloran}}]{madden2011}
{Madden}, S.~C., {Galametz}, M., {Cormier}, D., {et~al.} 2011, in EAS
  Publications Series, Vol.~52, EAS Publications Series, ed. {M.~R{\"o}llig,
  R.~Simon, V.~Ossenkopf, \& J.~Stutzki}, 95--101

\newblock {The Elusive ISM of Dwarf Galaxies: Excess Submillimetre Emission and
  CO-Dark Molecular Gas}.
\newblock In {M.~R{\"o}llig, R.~Simon, V.~Ossenkopf, \& J.~Stutzki} (Ed.), {\em
  EAS Publications Series}, Volume~52 of {\em EAS Publications Series}, pp.\
  95--101.

\bibitem[{{Marseille} {et~al.}(2010){Marseille}, {van der Tak}, {Herpin},
  {Wyrowski}, {Chavarr{\'{\i}}a}, {Pietropaoli}, {Baudry}, {Bontemps},
  {Cernicharo}, {Jacq}, {Frieswijk}, {Shipman}, {van Dishoeck}, {Bachiller},
  {Benedettini}, {Benz}, {Bergin}, {Bjerkeli}, {Blake}, {Braine}, {Bruderer},
  {Caselli}, {Caux}, {Codella}, {Daniel}, {Dieleman}, {di Giorgio}, {Dominik},
  {Doty}, {Encrenaz}, {Fich}, {Fuente}, {Gaier}, {Giannini}, {Goicoechea}, {de
  Graauw}, {Helmich}, {Herczeg}, {Hogerheijde}, {Jackson}, {Javadi}, {Jellema},
  {Johnstone}, {J{\o}rgensen}, {Kester}, {Kristensen}, {Larsson}, {Laauwen},
  {Lis}, {Liseau}, {Luinge}, {McCoey}, {Megej}, {Melnick}, {Neufeld}, {Nisini},
  {Olberg}, {Parise}, {Pearson}, {Plume}, {Risacher}, {Roelfsema},
  {Santiago-Garc{\'{\i}}a}, {Saraceno}, {Siegel}, {Stutzki}, {Tafalla}, {van
  Kempen}, {Visser}, {Wampfler}, \& {Y{\i}ld{\i}z}}]{marseille2010}
{Marseille}, M.~G., {van der Tak}, F.~F.~S., {Herpin}, F., {et~al.} 2010, \aap,
  521, L32

\newblock {Water abundances in high-mass protostellar envelopes: Herschel
  observations with HIFI}.
\newblock {\em \aap\/}~{\bf 521}, L32.

\bibitem[{{Mayor} {et~al.}(2011){Mayor}, {Marmier}, {Lovis}, {Udry},
  {S{\'e}gransan}, {Pepe}, {Benz}, {Bertaux}, {Bouchy}, {Dumusque}, {Lo Curto},
  {Mordasini}, {Queloz}, \& {Santos}}]{mayor2011}
{Mayor}, M., {Marmier}, M., {Lovis}, C., {et~al.} 2011, ArXiv 1109.2497

\newblock {The HARPS search for southern extra-solar planets XXXIV. Occurrence,
  mass distribution and orbital properties of super-Earths and Neptune-mass
  planets}.
\newblock {\em ArXiv 1109.2497\/}.

\bibitem[{{Melnick} \& {Bergin}(2005)}]{melnick2005}
{Melnick}, G.~J. \& {Bergin}, E.~A. 2005, Advances in Space Research, 36, 1027

\newblock {The legacy of SWAS: Water and molecular oxygen in the interstellar
  medium}.
\newblock {\em Advances in Space Research\/}~{\bf 36}, 1027--1030.

\bibitem[{{Melnick} {et~al.}(2001){Melnick}, {Neufeld}, {Ford}, {Hollenbach},
  \& {Ashby}}]{melnick2001}
{Melnick}, G.~J., {Neufeld}, D.~A., {Ford}, K.~E.~S., {Hollenbach}, D.~J., \&
  {Ashby}, M.~L.~N. 2001, \nat, 412, 160

\newblock {Discovery of water vapour around IRC+10216 as evidence for comets
  orbiting another star}.
\newblock {\em \nat\/}~{\bf 412}, 160--163.

\bibitem[{{Melnick} {et~al.}(2010){Melnick}, {Tolls}, {Neufeld}, {Bergin},
  {Phillips}, {Wang}, {Crockett}, {Bell}, {Blake}, {Cabrit}, {Caux},
  {Ceccarelli}, {Cernicharo}, {Comito}, {Daniel}, {Dubernet}, {Emprechtinger},
  {Encrenaz}, {Falgarone}, {Gerin}, {Giesen}, {Goicoechea}, {Goldsmith},
  {Herbst}, {Joblin}, {Johnstone}, {Langer}, {Latter}, {Lis}, {Lord}, {Maret},
  {Martin}, {Menten}, {Morris}, {M{\"u}ller}, {Murphy}, {Ossenkopf}, {Pagani},
  {Pearson}, {P{\'e}rault}, {Plume}, {Qin}, {Salez}, {Schilke}, {Schlemmer},
  {Stutzki}, {Trappe}, {van der Tak}, {Vastel}, {Yorke}, {Yu}, \&
  {Zmuidzinas}}]{melnick2010}
{Melnick}, G.~J., {Tolls}, V., {Neufeld}, D.~A., {et~al.} 2010, \aap, 521, L27

\newblock {Herschel observations of EXtra-Ordinary Sources (HEXOS):
  Observations of H$_{2}$O and its isotopologues towards Orion KL}.
\newblock {\em \aap\/}~{\bf 521}, L27.

\bibitem[{{Menten} {et~al.}(2010){Menten}, {Wyrowski}, {Alcolea}, {De Beck},
  {Decin}, {Marston}, {Bujarrabal}, {Cernicharo}, {Dominik}, {Justtanont}, {de
  Koter}, {Melnick}, {Neufeld}, {Olofsson}, {Planesas}, {Schmidt},
  {Sch{\"o}ier}, {Szczerba}, {Teyssier}, {Waters}, {Edwards}, {Olberg},
  {Phillips}, {Morris}, {Salez}, \& {Caux}}]{menten2010}
{Menten}, K.~M., {Wyrowski}, F., {Alcolea}, J., {et~al.} 2010, \aap, 521, L7

\newblock {Herschel/HIFI deepens the circumstellar NH$_{3}$ enigma}.
\newblock {\em \aap\/}~{\bf 521}, L7.

\bibitem[{{Monje} {et~al.}(2011{\natexlab{a}}){Monje}, {Emprechtinger},
  {Phillips}, {Lis}, {Goldsmith}, {Bergin}, {Bell}, {Neufeld}, \&
  {Sonnentrucker}}]{monje2011}
{Monje}, R.~R., {Emprechtinger}, M., {Phillips}, T.~G., {et~al.}
  2011{\natexlab{a}}, \apjl, 734, L23

\newblock {Herschel/HIFI Observations of Hydrogen Fluoride Toward Sagittarius
  B2(M)}.
\newblock {\em \apjl\/}~{\bf 734}, L23.

\bibitem[{{Monje} {et~al.}(2011{\natexlab{b}}){Monje}, {Phillips}, {Peng},
  {Lis}, {Neufeld}, \& {Emprechtinger}}]{monje2011clover}
{Monje}, R.~R., {Phillips}, T.~G., {Peng}, R., {et~al.} 2011{\natexlab{b}},
  \apjl, 742, L21

\newblock {Discovery of Hydrogen Fluoride in the Cloverleaf Quasar at z =
  2.56}.
\newblock {\em \apjl\/}~{\bf 742}, L21.

\bibitem[{{Mookerjea} {et~al.}(2011){Mookerjea}, {Kramer}, {Buchbender},
  {Boquien}, {Verley}, {Rela{\~n}o}, {Quintana-Lacaci}, {Aalto}, {Braine},
  {Calzetti}, {Combes}, {Garcia-Burillo}, {Gratier}, {Henkel}, {Israel},
  {Lord}, {Nikola}, {R{\"o}llig}, {Stacey}, {Tabatabaei}, {van der Tak}, \&
  {van der Werf}}]{mookerjea2011}
{Mookerjea}, B., {Kramer}, C., {Buchbender}, C., {et~al.} 2011, \aap, 532, A152

\newblock {The Herschel M 33 extended survey (HerM33es): PACS spectroscopy of
  the star-forming region BCLMP 302}.
\newblock {\em \aap\/}~{\bf 532}, A152.

\bibitem[{{Morbidelli}(2010)}]{morbidelli2010}
{Morbidelli}, A. 2010, Comptes Rendus Physique, 11, 651

\newblock {A coherent and comprehensive model of the evolution of the outer
  Solar System}.
\newblock {\em Comptes Rendus Physique\/}~{\bf 11}, 651--659.

\bibitem[{{Moreno} {et~al.}(2010){Moreno}, {Lellouch}, {Hartogh}, {Lara},
  {Courtin}, {Rengel}, {Jarchow}, {Bockel{\'e}e-Morvan}, {Biver}, {Lis}, \&
  {HssO Team}}]{moreno2010}
{Moreno}, R., {Lellouch}, E., {Hartogh}, P., {et~al.} 2010, in Bulletin of the
  American Astronomical Society, Vol.~42, AAS/Division for Planetary Sciences
  Meeting Abstracts \#42, 1088

\newblock {Herschel/HIFI Observations of Titan : Observation of the \hho\
  ($1_{10}-1_{01}$) 557 GHz Line and First Detection of HNC}.
\newblock In {\em AAS/Division for Planetary Sciences Meeting Abstracts \#42},
  Volume~42 of {\em Bulletin of the American Astronomical Society}, pp.\  1088.

\bibitem[{{Moreno} {et~al.}(2011){Moreno}, {Lellouch}, {Lara}, {Courtin},
  {Bockel{\'e}e-Morvan}, {Hartogh}, {Rengel}, {Biver}, {Banaszkiewicz}, \&
  {Gonz{\'a}lez}}]{moreno2011}
{Moreno}, R., {Lellouch}, E., {Lara}, L.~M., {et~al.} 2011, \aap, 536, L12

\newblock {First detection of hydrogen isocyanide (HNC) in Titan's atmosphere}.
\newblock {\em \aap\/}~{\bf 536}, L12.

\bibitem[{{Neufeld} {et~al.}(2010){Neufeld}, {Goicoechea}, {Sonnentrucker},
  {Black}, {Pearson}, {Yu}, {Phillips}, {Lis}, {de Luca}, {Herbst}, {Rimmer},
  {Gerin}, {Bell}, {Boulanger}, {Cernicharo}, {Coutens}, {Dartois},
  {Kazmierczak}, {Encrenaz}, {Falgarone}, {Geballe}, {Giesen}, {Godard},
  {Goldsmith}, {Gry}, {Gupta}, {Hennebelle}, {Hily-Blant}, {Joblin},
  {Ko{\l}os}, {Kre{\l}owski}, {Mart{\'{\i}}n-Pintado}, {Menten}, {Monje},
  {Mookerjea}, {Perault}, {Persson}, {Plume}, {Salez}, {Schlemmer}, {Schmidt},
  {Stutzki}, {Teyssier}, {Vastel}, {Cros}, {Klein}, {Lorenzani}, {Philipp},
  {Samoska}, {Shipman}, {Tielens}, {Szczerba}, \&
  {Zmuidzinas}}]{neufeld2010:hno+}
{Neufeld}, D.~A., {Goicoechea}, J.~R., {Sonnentrucker}, P., {et~al.} 2010,
  \aap, 521, L10

\newblock {Herschel/HIFI observations of interstellar OH$^{+}$ and
  H$_{2}$O$^{+}$ towards W49N: a probe of diffuse clouds with a small molecular
  fraction}.
\newblock {\em \aap\/}~{\bf 521}, L10.

\bibitem[{{Neufeld} {et~al.}(2011){Neufeld}, {Gonz{\'a}lez-Alfonso}, {Melnick},
  {Szczerba}, {Schmidt}, {Decin}, {de Koter}, {Sch{\"o}ier}, \&
  {Cernicharo}}]{neufeld2011}
{Neufeld}, D.~A., {Gonz{\'a}lez-Alfonso}, E., {Melnick}, G.~J., {et~al.} 2011,
  \apjl, 727, L28

\newblock {Herschel/HIFI Observations of IRC+10216: Water Vapor in the Inner
  Envelope of a Carbon-rich Asymptotic Giant Branch Star}.
\newblock {\em \apjl\/}~{\bf 727}, L28.

\bibitem[{{Neufeld} {et~al.}(1997){Neufeld}, {Zmuidzinas}, {Schilke}, \&
  {Phillips}}]{neufeld1997}
{Neufeld}, D.~A., {Zmuidzinas}, J., {Schilke}, P., \& {Phillips}, T.~G. 1997,
  \apjl, 488, L141

\newblock {Discovery of Interstellar Hydrogen Fluoride}.
\newblock {\em \apjl\/}~{\bf 488}, L141.

\bibitem[{Neufeld {et~al.}(2012)Neufeld, Snell, 4, 5, 6, 7, \& 8}]{neufeld2012}
Neufeld, D.~Roueff, E., Snell, R., 4, {et~al.} 2012, \apj, in press,
  arXiv:1201.2941

\newblock Herschel observations of interstellar chloronium.
\newblock {\em \apj\/}~{\bf in press, arXiv:1201.2941}.

\bibitem[{{Ossenkopf} {et~al.}(2010){Ossenkopf}, {M{\"u}ller}, {Lis},
  {Schilke}, {Bell}, {Bruderer}, {Bergin}, {Ceccarelli}, {Comito}, {Stutzki},
  {Bacman}, {Baudry}, {Benz}, {Benedettini}, {Berne}, {Blake}, {Boogert},
  {Bottinelli}, {Boulanger}, {Cabrit}, {Caselli}, {Caux}, {Cernicharo},
  {Codella}, {Coutens}, {Crimier}, {Crockett}, {Daniel}, {Demyk}, {Dieleman},
  {Dominik}, {Dubernet}, {Emprechtinger}, {Encrenaz}, {Falgarone}, {France},
  {Fuente}, {Gerin}, {Giesen}, {di Giorgio}, {Goicoechea}, {Goldsmith},
  {G{\"u}sten}, {Harris}, {Helmich}, {Herbst}, {Hily-Blant}, {Jacobs}, {Jacq},
  {Joblin}, {Johnstone}, {Kahane}, {Kama}, {Klein}, {Klotz}, {Kramer},
  {Langer}, {Lefloch}, {Leinz}, {Lorenzani}, {Lord}, {Maret}, {Martin},
  {Martin-Pintado}, {McCoey}, {Melchior}, {Melnick}, {Menten}, {Mookerjea},
  {Morris}, {Murphy}, {Neufeld}, {Nisini}, {Pacheco}, {Pagani}, {Parise},
  {Pearson}, {P{\'e}rault}, {Phillips}, {Plume}, {Quin}, {Rizzo}, {R{\"o}llig},
  {Salez}, {Saraceno}, {Schlemmer}, {Simon}, {Schuster}, {van der Tak},
  {Tielens}, {Teyssier}, {Trappe}, {Vastel}, {Viti}, {Wakelam}, {Walters},
  {Wang}, {Whyborn}, {van der Wiel}, {Yorke}, {Yu}, \&
  {Zmuidzinas}}]{ossenkopf2010}
{Ossenkopf}, V., {M{\"u}ller}, H.~S.~P., {Lis}, D.~C., {et~al.} 2010, \aap,
  518, L111

\newblock {Detection of interstellar oxidaniumyl: Abundant H$_{2}$O$^{+}$
  towards the star-forming regions DR21, Sgr B2, and NGC6334}.
\newblock {\em \aap\/}~{\bf 518}, L111.

\bibitem[{{Peng} {et~al.}(2010){Peng}, {Yoshida}, {Chamberlin}, {Phillips},
  {Lis}, \& {Gerin}}]{peng2010}
{Peng}, R., {Yoshida}, H., {Chamberlin}, R.~A., {et~al.} 2010, \apj, 723, 218

\newblock {A Comprehensive Survey of Hydrogen Chloride in the Galaxy}.
\newblock {\em \apj\/}~{\bf 723}, 218--228.

\bibitem[{{Perry} {et~al.}(2010){Perry}, {Teolis}, {Smith}, {McNutt},
  {Fletcher}, {Kasprzak}, {Magee}, {Mitchell}, \& {Waite}}]{perry2010}
{Perry}, M.~E., {Teolis}, B., {Smith}, H.~T., {et~al.} 2010, Journal of
  Geophysical Research (Space Physics), 115, 10206

\newblock {Cassini INMS observations of neutral molecules in Saturn's E-ring}.
\newblock {\em Journal of Geophysical Research (Space Physics)\/}~{\bf 115},
  10206.

\bibitem[{{Phillips} {et~al.}(2010){Phillips}, {Bergin}, {Lis}, {Neufeld},
  {Bell}, {Wang}, {Crockett}, {Emprechtinger}, {Blake}, {Caux}, {Ceccarelli},
  {Cernicharo}, {Comito}, {Daniel}, {Dubernet}, {Encrenaz}, {Gerin}, {Giesen},
  {Goicoechea}, {Goldsmith}, {Herbst}, {Joblin}, {Johnstone}, {Langer},
  {Latter}, {Lord}, {Maret}, {Martin}, {Melnick}, {Menten}, {Morris},
  {M{\"u}ller}, {Murphy}, {Ossenkopf}, {Pearson}, {P{\'e}rault}, {Plume},
  {Qin}, {Schilke}, {Schlemmer}, {Stutzki}, {Trappe}, {van der Tak}, {Vastel},
  {Yorke}, {Yu}, {Zmuidzinas}, {Boogert}, {G{\"u}sten}, {Hartogh}, {Honingh},
  {Karpov}, {Kooi}, {Krieg}, \& {Schieder}}]{phillips2010}
{Phillips}, T.~G., {Bergin}, E.~A., {Lis}, D.~C., {et~al.} 2010, \aap, 518,
  L109

\newblock {Herschel observations of EXtra-Ordinary Sources (HEXOS): Detection
  of hydrogen fluoride in absorption towards Orion KL}.
\newblock {\em \aap\/}~{\bf 518}, L109.

\bibitem[{{Pilbratt} {et~al.}(2010){Pilbratt}, {Riedinger}, {Passvogel},
  {Crone}, {Doyle}, {Gageur}, {Heras}, {Jewell}, {Metcalfe}, {Ott}, \&
  {Schmidt}}]{pilbratt2010}
{Pilbratt}, G.~L., {Riedinger}, J.~R., {Passvogel}, T., {et~al.} 2010, \aap,
  518, L1

\newblock {Herschel Space Observatory. An ESA facility for far-infrared and
  submillimetre astronomy}.
\newblock {\em \aap\/}~{\bf 518}, L1.

\bibitem[{{Plume} {et~al.}(2012){Plume}, {Bergin}, {Phillips}, {Lis}, {Wang},
  {Crockett}, {Caux}, {Comito}, {Goldsmith}, \& {Schilke}}]{plume2012}
{Plume}, R., {Bergin}, E.~A., {Phillips}, T.~G., {et~al.} 2012, \apj, 744, 28

\newblock {A Direct Measurement of the Total Gas Column Density in Orion KL}.
\newblock {\em \apj\/}~{\bf 744}, 28.

\bibitem[{{Rangwala} {et~al.}(2011){Rangwala}, {Maloney}, {Glenn}, {Wilson},
  {Rykala}, {Isaak}, {Baes}, {Bendo}, {Boselli}, {Bradford}, {Clements},
  {Cooray}, {Fulton}, {Imhof}, {Kamenetzky}, {Madden}, {Mentuch}, {Sacchi},
  {Sauvage}, {Schirm}, {Smith}, {Spinoglio}, \& {Wolfire}}]{rangwala2011}
{Rangwala}, N., {Maloney}, P.~R., {Glenn}, J., {et~al.} 2011, \apj, 743, 94

\newblock {Observations of Arp 220 Using Herschel-SPIRE: An Unprecedented View
  of the Molecular Gas in an Extreme Star Formation Environment}.
\newblock {\em \apj\/}~{\bf 743}, 94.

\bibitem[{{Roelfsema} {et~al.}(2012){Roelfsema}, {Helmich}, {Teyssier},
  {Ossenkopf}, {Morris}, {Olberg}, {Shipman}, {Risacher}, {Akyilmaz},
  {Assendorp}, {Avruch}, {Beintema}, {Biver}, {Boogert}, {Borys}, {Braine},
  {Caris}, {Caux}, {Cernicharo}, {Coeur-Joly}, {Comito}, {de Lange},
  {Delforge}, {Dieleman}, {Dubbeldam}, {de Graauw}, {Edwards}, {Fich},
  {Flederus}, {Gal}, {di Giorgio}, {Herpin}, {Higgins}, {Hoac}, {Huisman},
  {Jarchow}, {Jellema}, {de Jonge}, {Kester}, {Klein}, {Kooi}, {Kramer},
  {Laauwen}, {Larsson}, {Leinz}, {Lord}, {Lorenzani}, {Luinge}, {Marston},
  {Mart{\'{\i}}n-Pintado}, {McCoey}, {Melchior}, {Michalska}, {Moreno},
  {M{\"u}ller}, {Nowosielski}, {Okada}, {Orlea{\'n}ski}, {Phillips}, {Pearson},
  {Rabois}, {Ravera}, {Rector}, {Rengel}, {Sagawa}, {Salomons},
  {S{\'a}nchez-Su{\'a}rez}, {Schieder}, {Schl{\"o}der}, {Schm{\"u}lling},
  {Soldati}, {Stutzki}, {Thomas}, {Tielens}, {Vastel}, {Wildeman}, {Xie},
  {Xilouris}, {Wafelbakker}, {Whyborn}, {Zaal}, {Bell}, {Bjerkeli}, {De Beck},
  {Cavali{\'e}}, {Crockett}, {Hily-Blant}, {Kama}, {Kaminski}, {Lefl{\'o}ch},
  {Lombaert}, {de Luca}, {Makai}, {Marseille}, {Nagy}, {Pacheco}, {van der
  Wiel}, {Wang}, \& {Y{\i}ld{\i}z}}]{roelfsema2012}
{Roelfsema}, P.~R., {Helmich}, F.~P., {Teyssier}, D., {et~al.} 2012, \aap, 537,
  A17

\newblock {In-orbit performance of Herschel-HIFI}.
\newblock {\em \aap\/}~{\bf 537}, A17.

\bibitem[{{Snow} \& {McCall}(2006)}]{snow2006}
{Snow}, T.~P. \& {McCall}, B.~J. 2006, \araa, 44, 367

\newblock {Diffuse Atomic and Molecular Clouds}.
\newblock {\em \araa\/}~{\bf 44}, 367--414.

\bibitem[{{Sonnentrucker} {et~al.}(2010){Sonnentrucker}, {Neufeld}, {Phillips},
  {Gerin}, {Lis}, {de Luca}, {Goicoechea}, {Black}, {Bell}, {Boulanger},
  {Cernicharo}, {Coutens}, {Dartois}, {Ka{\'z}mierczak}, {Encrenaz},
  {Falgarone}, {Geballe}, {Giesen}, {Godard}, {Goldsmith}, {Gry}, {Gupta},
  {Hennebelle}, {Herbst}, {Hily-Blant}, {Joblin}, {Ko{\l}os}, {Kre{\l}owski},
  {Mart{\'{\i}}n-Pintado}, {Menten}, {Monje}, {Mookerjea}, {Pearson},
  {Perault}, {Persson}, {Plume}, {Salez}, {Schlemmer}, {Schmidt}, {Stutzki},
  {Teyssier}, {Vastel}, {Yu}, {Caux}, {G{\"u}sten}, {Hatch}, {Klein}, {Mehdi},
  {Morris}, \& {Ward}}]{sonnentrucker2010}
{Sonnentrucker}, P., {Neufeld}, D.~A., {Phillips}, T.~G., {et~al.} 2010, \aap,
  521, L12

\newblock {Detection of hydrogen fluoride absorption in diffuse molecular
  clouds with Herschel/HIFI: an ubiquitous tracer of molecular gas}.
\newblock {\em \aap\/}~{\bf 521}, L12.

\bibitem[{{Stahler} \& {Palla}(2005)}]{palla-stahler}
{Stahler}, S.~W. \& {Palla}, F. 2005, {The Formation of Stars} (Wiley-VCH)

\newblock {\em {The Formation of Stars}}.
\newblock Wiley-VCH.

\bibitem[{Van~der Tak(2012)}]{vandertak2012:h3+}
Van~der Tak, F.~F.~S. 2012, Phil. Trans. R. Soc. A, submitted

\newblock Using deuterated \hhhp\ and other molecular ions to probe the
  formation of stars and planets.
\newblock {\em Phil. Trans. R. Soc. A\/}~{\bf submitted}.

\bibitem[{{Van der Tak} {et~al.}(2010){Van der Tak}, {Marseille}, {Herpin},
  {Wyrowski}, {Baudry}, {Bontemps}, {Braine}, {Doty}, {Frieswijk}, {Melnick},
  {Shipman}, {van Dishoeck}, {Benz}, {Caselli}, {Hogerheijde}, {Johnstone},
  {Liseau}, {Bachiller}, {Benedettini}, {Bergin}, {Bjerkeli}, {Blake},
  {Bruderer}, {Cernicharo}, {Codella}, {Daniel}, {di Giorgio}, {Dominik},
  {Encrenaz}, {Fich}, {Fuente}, {Giannini}, {Goicoechea}, {de Graauw},
  {Helmich}, {Herczeg}, {J{\o}rgensen}, {Kristensen}, {Larsson}, {Lis},
  {McCoey}, {Neufeld}, {Nisini}, {Olberg}, {Parise}, {Pearson}, {Plume},
  {Risacher}, {Santiago}, {Saraceno}, {Tafalla}, {van Kempen}, {Visser},
  {Wampfler}, {Y{\i}ld{\i}z}, {Ravera}, {Roelfsema}, {Siebertz}, \&
  {Teyssier}}]{vandertak2010}
{Van der Tak}, F.~F.~S., {Marseille}, M.~G., {Herpin}, F., {et~al.} 2010, \aap,
  518, L107

\newblock {Water abundance variations around high-mass protostars: HIFI
  observations of the DR21 region}.
\newblock {\em \aap\/}~{\bf 518}, L107.

\bibitem[{{Van der Tak} {et~al.}(2012){Van der Tak}, {Ossenkopf}, {Nagy},
  {Faure}, {R{\"o}llig}, \& {Bergin}}]{vandertak2012:hf}
{Van der Tak}, F.~F.~S., {Ossenkopf}, V., {Nagy}, Z., {et~al.} 2012, \aap, 537,
  L10

\newblock {Detection of HF emission from the Orion Bar}.
\newblock {\em \aap\/}~{\bf 537}, L10.

\bibitem[{{Van der Werf} {et~al.}(2010){Van der Werf}, {Isaak}, {Meijerink},
  {Spaans}, {Rykala}, {Fulton}, {Loenen}, {Walter}, {Wei{\ss}}, {Armus},
  {Fischer}, {Israel}, {Harris}, {Veilleux}, {Henkel}, {Savini}, {Lord},
  {Smith}, {Gonz{\'a}lez-Alfonso}, {Naylor}, {Aalto}, {Charmandaris}, {Dasyra},
  {Evans}, {Gao}, {Greve}, {G{\"u}sten}, {Kramer}, {Mart{\'{\i}}n-Pintado},
  {Mazzarella}, {Papadopoulos}, {Sanders}, {Spinoglio}, {Stacey}, {Vlahakis},
  {Wiedner}, \& {Xilouris}}]{vanderwerf2010}
{Van der Werf}, P.~P., {Isaak}, K.~G., {Meijerink}, R., {et~al.} 2010, \aap,
  518, L42

\newblock {Black hole accretion and star formation as drivers of gas excitation
  and chemistry in Markarian 231}.
\newblock {\em \aap\/}~{\bf 518}, L42.

\bibitem[{{Van Dishoeck} {et~al.}(2011){Van Dishoeck}, {Kristensen}, {Benz},
  {Bergin}, {Caselli}, {Cernicharo}, {Herpin}, {Hogerheijde}, {Johnstone},
  {Liseau}, {Nisini}, {Shipman}, {Tafalla}, {van der Tak}, {Wyrowski},
  {Aikawa}, {Bachiller}, {Baudry}, {Benedettini}, {Bjerkeli}, {Blake},
  {Bontemps}, {Braine}, {Brinch}, {Bruderer}, {Chavarr{\'{\i}}a}, {Codella},
  {Daniel}, {de Graauw}, {Deul}, {di Giorgio}, {Dominik}, {Doty}, {Dubernet},
  {Encrenaz}, {Feuchtgruber}, {Fich}, {Frieswijk}, {Fuente}, {Giannini},
  {Goicoechea}, {Helmich}, {Herczeg}, {Jacq}, {J{\o}rgensen}, {Karska},
  {Kaufman}, {Keto}, {Larsson}, {Lefloch}, {Lis}, {Marseille}, {McCoey},
  {Melnick}, {Neufeld}, {Olberg}, {Pagani}, {Pani{\'c}}, {Parise}, {Pearson},
  {Plume}, {Risacher}, {Salter}, {Santiago-Garc{\'{\i}}a}, {Saraceno},
  {St{\"a}uber}, {van Kempen}, {Visser}, {Viti}, {Walmsley}, {Wampfler}, \&
  {Y{\i}ld{\i}z}}]{vandishoeck2011}
{Van Dishoeck}, E.~F., {Kristensen}, L.~E., {Benz}, A.~O., {et~al.} 2011,
  \pasp, 123, 138

\newblock {Water in Star-forming Regions with the Herschel Space Observatory
  (WISH). I. Overview of Key Program and First Results}.
\newblock {\em \pasp\/}~{\bf 123}, 138--170.

\bibitem[{{Velusamy} {et~al.}(2010){Velusamy}, {Langer}, {Pineda}, {Goldsmith},
  {Li}, \& {Yorke}}]{velusamy2010}
{Velusamy}, T., {Langer}, W.~D., {Pineda}, J.~L., {et~al.} 2010, \aap, 521, L18

\newblock {[CII] observations of H$_{2}$ molecular layers in transition
  clouds}.
\newblock {\em \aap\/}~{\bf 521}, L18.

\bibitem[{{Visser} {et~al.}(2009){Visser}, {van Dishoeck}, \&
  {Black}}]{visser2009}
{Visser}, R., {van Dishoeck}, E.~F., \& {Black}, J.~H. 2009, \aap, 503, 323

\newblock {The photodissociation and chemistry of CO isotopologues:
  applications to interstellar clouds and circumstellar disks}.
\newblock {\em \aap\/}~{\bf 503}, 323--343.

\bibitem[{{Wampfler} {et~al.}(2011){Wampfler}, {Bruderer}, {Kristensen},
  {Chavarr{\'{\i}}a}, {Bergin}, {Benz}, {van Dishoeck}, {Herczeg}, {van der
  Tak}, {Goicoechea}, {Doty}, \& {Herpin}}]{wampfler2011}
{Wampfler}, S.~F., {Bruderer}, S., {Kristensen}, L.~E., {et~al.} 2011, \aap,
  531, L16

\newblock {First hyperfine resolved far-infrared OH spectrum from a
  star-forming region}.
\newblock {\em \aap\/}~{\bf 531}, L16.

\bibitem[{{Wang} {et~al.}(2011){Wang}, {Bergin}, {Crockett}, {Goldsmith},
  {Lis}, {Pearson}, {Schilke}, {Bell}, {Comito}, {Blake}, {Caux}, {Ceccarelli},
  {Cernicharo}, {Daniel}, {Dubernet}, {Emprechtinger}, {Encrenaz}, {Gerin},
  {Giesen}, {Goicoechea}, {Gupta}, {Herbst}, {Joblin}, {Johnstone}, {Langer},
  {Latter}, {Lord}, {Maret}, {Martin}, {Melnick}, {Menten}, {Morris},
  {M{\"u}ller}, {Murphy}, {Neufeld}, {Ossenkopf}, {P{\'e}rault}, {Phillips},
  {Plume}, {Qin}, {Schlemmer}, {Stutzki}, {Trappe}, {van der Tak}, {Vastel},
  {Yorke}, {Yu}, \& {Zmuidzinas}}]{wang2011}
{Wang}, S., {Bergin}, E.~A., {Crockett}, N.~R., {et~al.} 2011, \aap, 527, A95

\newblock {Herschel observations of EXtra-Ordinary Sources (HEXOS): Methanol as
  a probe of physical conditions in Orion KL}.
\newblock {\em \aap\/}~{\bf 527}, A95.

\bibitem[{{Wei{\ss}} {et~al.}(2010){Wei{\ss}}, {Requena-Torres}, {G{\"u}sten},
  {Garc{\'{\i}}a-Burillo}, {Harris}, {Israel}, {Klein}, {Kramer}, {Lord},
  {Martin-Pintado}, {R{\"o}llig}, {Stutzki}, {Szczerba}, {van der Werf},
  {Philipp-May}, {Yorke}, {Akyilmaz}, {Gal}, {Higgins}, {Marston}, {Roberts},
  {Schl{\"o}der}, {Schultz}, {Teyssier}, {Whyborn}, \& {Wunsch}}]{weiss2010}
{Wei{\ss}}, A., {Requena-Torres}, M.~A., {G{\"u}sten}, R., {et~al.} 2010, \aap,
  521, L1

\newblock {HIFI spectroscopy of low-level water transitions in M 82}.
\newblock {\em \aap\/}~{\bf 521}, L1.

\bibitem[{{Whittet}(2010)}]{whittet2010}
{Whittet}, D.~C.~B. 2010, \apj, 710, 1009

\newblock {Oxygen Depletion in the Interstellar Medium: Implications for Grain
  Models and the Distribution of Elemental Oxygen}.
\newblock {\em \apj\/}~{\bf 710}, 1009--1016.

\bibitem[{{Willner} {et~al.}(1982){Willner}, {Gillett}, {Herter}, {Jones},
  {Krassner}, {Merrill}, {Pipher}, {Puetter}, {Rudy}, {Russell}, \&
  {Soifer}}]{willner1982}
{Willner}, S.~P., {Gillett}, F.~C., {Herter}, T.~L., {et~al.} 1982, \apj, 253,
  174

\newblock {Infrared spectra of protostars - Composition of the dust shells}.
\newblock {\em \apj\/}~{\bf 253}, 174--187.

\bibitem[{{Wyrowski} {et~al.}(2010){Wyrowski}, {van der Tak}, {Herpin},
  {Baudry}, {Bontemps}, {Chavarria}, {Frieswijk}, {Jacq}, {Marseille},
  {Shipman}, {van Dishoeck}, {Benz}, {Caselli}, {Hogerheijde}, {Johnstone},
  {Liseau}, {Bachiller}, {Benedettini}, {Bergin}, {Bjerkeli}, {Blake},
  {Braine}, {Bruderer}, {Cernicharo}, {Codella}, {Daniel}, {di Giorgio},
  {Dominik}, {Doty}, {Encrenaz}, {Fich}, {Fuente}, {Giannini}, {Goicoechea},
  {de Graauw}, {Helmich}, {Herczeg}, {J{\o}rgensen}, {Kristensen}, {Larsson},
  {Lis}, {McCoey}, {Melnick}, {Nisini}, {Olberg}, {Parise}, {Pearson}, {Plume},
  {Risacher}, {Santiago}, {Saraceno}, {Tafalla}, {van Kempen}, {Visser},
  {Wampfler}, {Y{\i}ld{\i}z}, {Black}, {Falgarone}, {Gerin}, {Roelfsema},
  {Dieleman}, {Beintema}, {de Jonge}, {Whyborn}, {Stutzki}, \&
  {Ossenkopf}}]{wyrowski2010}
{Wyrowski}, F., {van der Tak}, F., {Herpin}, F., {et~al.} 2010, \aap, 521, L34

\newblock {Variations in H$_{2}$O$^{+}$/H$_{2}$O ratios toward massive
  star-forming regions}.
\newblock {\em \aap\/}~{\bf 521}, L34.

\bibitem[{{Y{\i}ld{\i}z} {et~al.}(2010){Y{\i}ld{\i}z}, {van Dishoeck},
  {Kristensen}, {Visser}, {J{\o}rgensen}, {Herczeg}, {van Kempen},
  {Hogerheijde}, {Doty}, {Benz}, {Bruderer}, {Wampfler}, {Deul}, {Bachiller},
  {Baudry}, {Benedettini}, {Bergin}, {Bjerkeli}, {Blake}, {Bontemps}, {Braine},
  {Caselli}, {Cernicharo}, {Codella}, {Daniel}, {di Giorgio}, {Dominik},
  {Encrenaz}, {Fich}, {Fuente}, {Giannini}, {Goicoechea}, {de Graauw},
  {Helmich}, {Herpin}, {Jacq}, {Johnstone}, {Larsson}, {Lis}, {Liseau}, {Liu},
  {Marseille}, {McCoey}, {Melnick}, {Neufeld}, {Nisini}, {Olberg}, {Parise},
  {Pearson}, {Plume}, {Risacher}, {Santiago-Garc{\'{\i}}a}, {Saraceno},
  {Shipman}, {Tafalla}, {Tielens}, {van der Tak}, {Wyrowski}, {Dieleman},
  {Jellema}, {Ossenkopf}, {Schieder}, \& {Stutzki}}]{yildiz2010}
{Y{\i}ld{\i}z}, U.~A., {van Dishoeck}, E.~F., {Kristensen}, L.~E., {et~al.}
  2010, \aap, 521, L40

\end{thebibliography}


\end{document}